\documentclass[
 reprint,amsmath,amssymb,aps,pra,
]{revtex4-2}
\usepackage{graphicx} 
\usepackage{multibib}
\newcites{methods}{References}
\setcitestyle{super}

\begin{document}

\title{
Scalable microwave-to-optical transducers at single photon level with spins
}

\author{Tian Xie}
    \thanks{These authors contributed equally to this work.}
	\affiliation{Kavli Nanoscience Institute and Thomas J. Watson, Sr., Laboratory of Applied Physics, California Institute of Technology, Pasadena, California 91125, USA}
	\affiliation{Institute for Quantum Information and Matter, California Institute of Technology, Pasadena, California 91125, USA}
\author{Rikuto Fukumori}
    \thanks{These authors contributed equally to this work.}
	\affiliation{Kavli Nanoscience Institute and Thomas J. Watson, Sr., Laboratory of Applied Physics, California Institute of Technology, Pasadena, California 91125, USA}
	\affiliation{Institute for Quantum Information and Matter, California Institute of Technology, Pasadena, California 91125, USA}
 \author{Jiahui Li}
	\affiliation{Kavli Nanoscience Institute and Thomas J. Watson, Sr., Laboratory of Applied Physics, California Institute of Technology, Pasadena, California 91125, USA}
	\affiliation{Institute for Quantum Information and Matter, California Institute of Technology, Pasadena, California 91125, USA}
\author{Andrei Faraon}
    \email[]{faraon@caltech.edu}
	\affiliation{Kavli Nanoscience Institute and Thomas J. Watson, Sr., Laboratory of Applied Physics, California Institute of Technology, Pasadena, California 91125, USA}
	\affiliation{Institute for Quantum Information and Matter, California Institute of Technology, Pasadena, California 91125, USA}

\begin{abstract}
Microwave-to-optical transduction of single photons will play an essential role in interconnecting future superconducting quantum devices, with applications in distributed quantum computing and secure communications. Various transducers that couple microwave and optical modes via an optical drive have been developed, utilizing nonlinear phenomena such as the Pockels effect and a combination of electromechanical, piezoelectric, and optomechanical couplings. However, the limited strength of these nonlinearities, set by bulk material properties, requires the use of high quality factor resonators, often in conjunction with sophisticated nano-fabrication of suspended structures. Thus, an efficient and scalable transduction technology is still an outstanding goal. Rare-earth ion (REI) doped crystals provide high-quality atomic resonances that result in effective second-order nonlinearities stronger by many orders of magnitude compared to conventional materials. Here, we use ytterbium-171 ions doped in a YVO$_4$ crystal at 340~ppm with an effective resonant $\chi^{(2)}$ nonlinearity of $\sim 10^7$ pm/V to implement an on-chip microwave-to-optical transducer. Without an engineered optical cavity, we achieve percent-level efficiencies with an added noise as low as 1.24(9) photons. To showcase scalability, we demonstrate the interference of photons originating from two simultaneously operated transducers, enabled by the inherent absolute frequencies of the atomic transitions. These results establish REI-based transducers as a highly competitive transduction platform, provide existing REI-based quantum technologies a native link to various leading quantum microwave platforms, and pave the way toward remote transducer-assisted entanglement of superconducting quantum machines.
\end{abstract}

\maketitle

\section{Introduction}
Quantum networks\cite{kimble2008quantum} for establishing entanglement at long distances will enable applications in quantum computing \cite{cirac1999distributed,arute2019quantum}, communications\cite{lo2014secure,pompili2021realization}, and sensing\cite{baumgratz2016quantum, Pirandola2018}. Given recent advances in superconducting quantum computing\cite{google2023suppressing,kim2023evidence}, developing technologies for remote entanglement between these machines will unlock new opportunities where all processing is done in the microwave domain and communications are done optically. A transducer that converts photons between microwave and optical frequencies is a key component inside such a hybrid network. The desired performance metrics of a transducer are high efficiency, low added noise, large bandwidth, and high repetition rate\cite{lauk2020perspectives}. Robustness and scalability are also paramount for future adoption in large-scale quantum networks.

To realize this challenging task, establishing efficient coupling between microwave and optical fields is imperative \cite{han2021microwave}. Various material platforms have demonstrated conversion, either through direct coupling via the $\chi^{(2)}$ nonlinearity in electro-optics \cite{sahu2022quantum,xu2021bidirectional,shen2024photonic}, or by utilizing a combination of coupling mechanisms such as piezo-electricity, electromechanics, optomechanics \cite{jiang2020efficient,weaver2024integrated,higginbotham2018harnessing,zhao2024quantum}, and atomic ensembles \cite{kumar2023quantum,rochman2023microwave,fernandez2019cavity}. With developed transducers, optical readout of a superconducting qubit\cite{mirhosseini2020superconducting,delaney2022superconducting,arnold2023all} and generation of entangled microwave-optical photon pairs \cite{sahu2023entangling,meesala2024non,meesala2023quantum} have been demonstrated. While significant progress has been made, no single platform has met all of the aforementioned metrics, thereby driving ongoing exploration and innovation across various approaches. In particular, the intrinsically weak nonlinearities of conventional materials \cite{hamze2020design} necessitate the use of high quality factor resonators, presenting difficulties in scaling and frequency matching, partly due to large variations in the operating frequencies from fabrication disorders. This has limited previous demonstrations to operating with a single transducer.

Rare-earth ion (REI)-doped crystals have been demonstrated as a promising platform for transduction\cite{williamson2014magneto} due to their highly coherent and narrow atomic transitions in both microwave and optical domains\cite{zhong2015optically,kindem2018characterization}. These atomic properties contribute to an ultra-strong effective nonlinearity between microwave and optical fields, significantly enhancing the transduction efficiency with simple device designs. Moreover, on-chip integration with REI-doped crystals has been successfully demonstrated \cite{bartholomew2020chip,zhou2023photonic}, paving the way for scalability. Furthermore, utilizing atomic ensembles provides an absolute frequency reference, thereby naturally addressing the challenge of frequency matching in remote entanglement tasks\cite{hatipoglu2024situ}. Lastly, REI-based quantum technologies have flourished across multiple areas including single-photon sources \cite{kindem2020control,ourari2023indistinguishable}, quantum memories \cite{hedges2010efficient,lago2021telecom}, and entanglement generation\cite{ruskuc2024scalable}. A necessary demand arises for a seamless link between various REI-based technologies and other leading microwave quantum platforms.
\begin{figure*}
\centering
\includegraphics[width=0.95\linewidth]{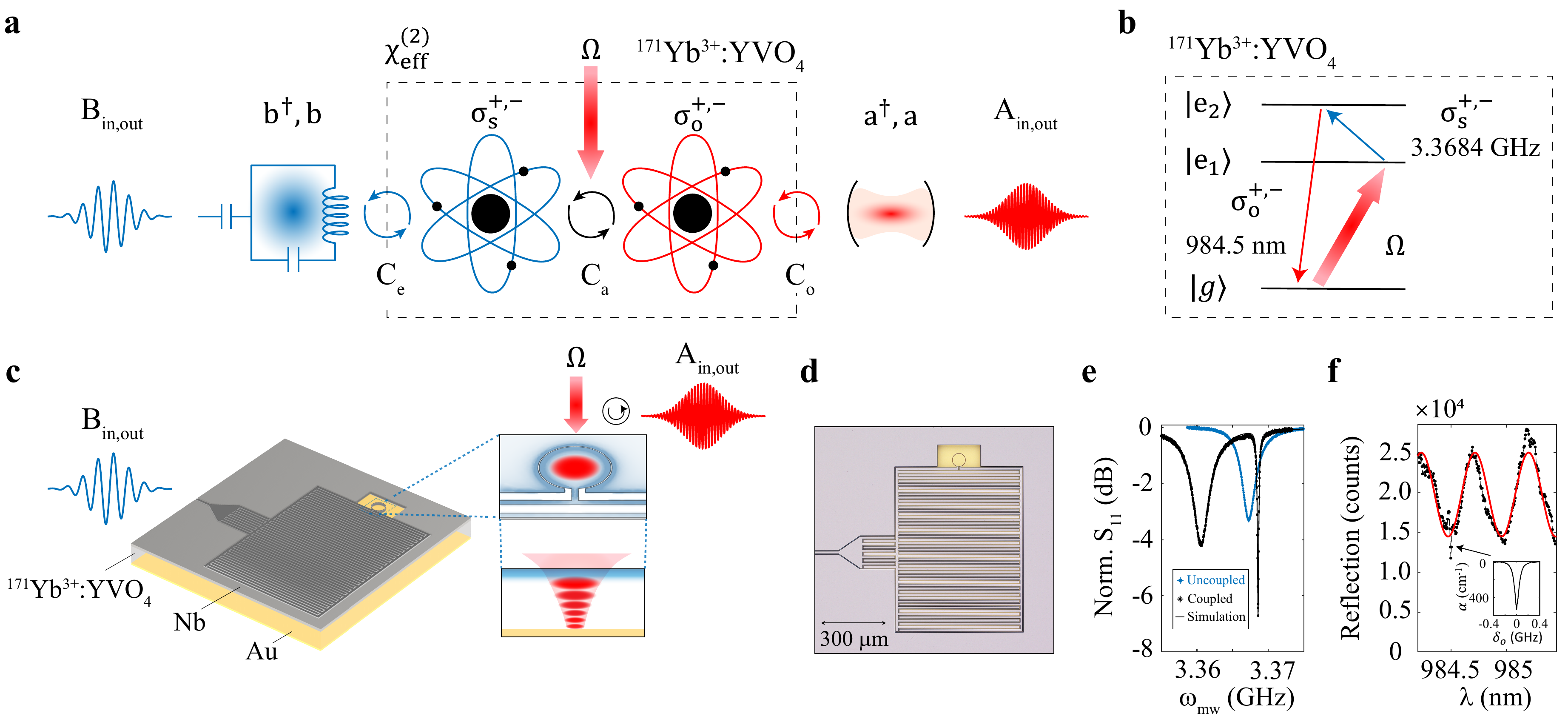}
\caption{\label{Fig1}
Concept and implementation of a REI-based on-chip microwave-to-optical transducer. (a) Concept of REI-based transducer. $B_{in,out}$ and $A_{in,out}$ are the input/output microwave and optical modes, $b^{\dagger},b$ and $a^{\dagger},a$ are the microwave and optical cavity modes, $C_e$, $C_a$, and $C_o$ are microwave, atomic and optical cooperativities, $\sigma_s^{+,-}$ and $\sigma_o^{+,-}$ are the spin and optical coherence operators. The black dashed box indicates the atomic part of the transduction process, where we define a material-dependent parameter $\chi^{(2)}_{eff}$. (b) The relevant energy levels of $^{171}$Yb$^{3+}:$YVO$_{4}$, forming a V-system for transduction. (c) Transducer device schematic. Microwave photons are sent in and emitted out via a coplanar waveguide and coupling capacitor. The magnetic energy for coupling to spins is confined within the circular inductor, where the optical pump is focused through the center of the circle. A weak optical mode is formed between the front interface and the back gold mirror. (d) Optical image of a fabricated device. (e) Microwave resonator and ion-cavity coupling spectra. The blue curve shows the uncoupled microwave resonator with no pump light. The black curve shows the hybridized ion-cavity coupling, obtained by turning the optical pump on and transferring population to $|e_1\rangle$. The narrow dip corresponds to the spin and the broad dip to the resonator. (f) Optical mode and ion inhomogeneity spectra. The oscillations show a period of $\sim0.5$~nm, characteristic of a weak Fabry-Perot mode formed between the gold back mirror and the top surface of the $500$~$\mu$m thick substrate. The inset shows a zoomed-in absorption profile of the $|g\rangle$-$|e_1\rangle$ transition.
}
\end{figure*}

In this work, we present a REI-based on-chip microwave-to-optical transducer, achieving a percent-level efficiency without an engineered optical cavity. We resolve the transduction of a single classical microwave photon to an optical photon on a single-photon detector, observing an added noise approaching quantum-enabled operation. Furthermore, we measure the interference of converted optical photons coming from two simultaneously operated transducers, demonstrating the benefit of an atomic-ensemble-based transducer in the context of scalability and its potential for facilitating remote entanglement.

\section{Results}
\subsection{Transducer concept and implementation}
Microwave-to-optical (M2O) transduction using a resonant three-level system in REIs can be visualized as a three-stage process, where each stage bridges two different modes with a coupling mechanism (Fig.~1a,b). Specifically, a microwave photon is first converted to spin coherence via microwave resonator coupling, then to optical coherence via an optical pump, and finally to an optical photon via coupling to an optical mode usually confined in an optical resonator (in reverse for optical-to-microwave (O2M)). Each stage can be parameterized by cooperativity, a dimensionless figure of merit comparing the coupling strength $g$ to the combined loss and decoherence across the two modes ($\kappa_1$ and $\kappa_2$) as $C= 4g^2/(\kappa_1\kappa_2)$. To gain physical intuition and capture the key physics of the conversion process, we use these cooperativities and derive an approximate end-to-end photon conversion efficiency $\eta$ as (Supplementary Information):
\begin{equation}
    \eta  \approx  r_e r_o \frac{C_e}{1+C_e} \frac{4C_a'}{(1+C_a')^2} \frac{C_o}{1+C_o}.
    \label{eq-eff}
\end{equation}
Here $C_e$ and $C_o$ are the microwave and optical cooperativities, $r_{e,o}$ are the extraction factors from the microwave and optical cavities, and we define the modified atomic cooperativity $C_a'= \frac{C_a}{(1+C_e)(1+C_o)}$ where $C_a$ is the atomic cooperativity without cavity coupling (see Methods for exact expressions). Unit efficiency can be achieved if $C_a'=1$ and $C_e,C_o>>1$, motivating the need for both high coupling strengths and low system decoherence. While to some extent the cavities can be engineered to have low decay rates, the intrinsic material properties hold considerable significance in maximizing the cooperativities.

These material properties can be encapsulated as the nonlinearity, for instance, the $\chi^{(2)}$ nonlinearity in electro-optics. Similarly for an atomic system with three relevant levels and transitions, we derive an effective non-linearity $\chi^{(2)}_{eff}$ in the low-cooperativity regime as (Supplementary Information):
\begin{equation}
    \chi_{eff}^{(2)} = \frac{4}{\varepsilon_0 c h^2}\frac{\rho d_{p} d_{o} \mu}{\Gamma_e \Gamma_o},
\end{equation}
where $\epsilon_0$ is the vacuum permittivity, $c$ is the speed of light, and $h$ is Planck's constant. This result indicates that strong optical dipole moments on the pump and signal arms ($d_p$ and $d_o$), strong spin dipole moment ($\mu$), and high concentration ($\rho$) are beneficial, whereas the ensemble decoherence (inhomogeneous linewidths $\Gamma_{o,e}$) is detrimental for conversion. As $\eta$ scales quadratically with $\chi^{(2)}$ in the low-cooperativity regime (Supplementary Information), it is desirable to have a large $\chi_{eff}^{(2)}$.

To this end, we use an ensemble of ytterbium-171 ions doped at $340$~ppm in yttrium-orthovanadate ($^{171}$Yb$^{3+}$:YVO$_4$), which has been shown to have strong dipole moments and narrow inhomogeneities, even at relatively high doping concentrations \cite{kindem2018characterization}. The relevant energy levels are shown in Fig.~1b, where we measure optical inhomogeneous linewidths of $\Gamma_o=2\pi\times 92(1)$~MHz for both the pump and the signal transitions and $\Gamma_e=2\pi\times 160(5)$~kHz for the spin transition. We note that the spin linewidth is substantially narrower than in other REI systems \cite{rochman2023microwave,probst2013anisotropic} due to the first-order insensitivity to magnetic fields of this hybridized electro-nuclear spin transition. With these numbers, we calculate $\chi_{eff}^{(2)} = 2 \times 10^7$ pm/V, over 4 orders of magnitude larger than a common electro-optic material LiNbO$_3$\cite{hamze2020design} with $n^3r_{51} =$ 400~pm/V.

To couple the ions to both microwave and optical modes, we utilize an on-chip superconducting microwave resonator and a low finesse free-space optical Fabry-Perot mode (Fig.~1c). The microwave resonator consists of an interdigitated capacitor and a circular inductor with a $30$~$\mu$m radius. It is coupled to an input/output port via a coupling capacitor and a coplanar waveguide. A low finesse optical mode is formed between the top (air-YVO$_4$ interface) and bottom (evaporated gold layer) surfaces. We note that the spin transition is in the optical excited state, such that only the spins that experience the pump participate in the transduction process. Therefore, by focusing the pump within the transduction zone defined by the circular inductor, we minimize the parasitic spins that only absorb microwave photons but do not contribute to the conversion process.

The microwave cavity has a measured decay rate $\kappa_{e}=2\pi\times 3$~MHz with no pump light, which was deliberately designed to be broad enough to interface with most superconducting qubits \cite{kjaergaard2020superconducting}. With the pump light turned on, some population is transferred to $|e_1\rangle$ and coupling to the microwave resonator is observed (Fig.~1e). We fit the data with a theoretical model (Supplementary Information) and extract the ensemble microwave cooperativity $C_e= 2.3$ (Methods). On the optical side, we observe a weak interference pattern that corresponds to the chip thickness (Fig.~1f). The low finesse ($\mathcal{F}=1.6$) indicates that most of the light goes through the crystal once and then leaks out, during which the transduction process happens with $C_o = 0.14$ (Supplementary Information). There is also a sharp absorption dip on top of the interference pattern, which corresponds to the optical transitions. A full optical absorption spectrum including both the pump and signal transitions are shown in Extended Data Fig.~1, and all experimental parameters are summarized in Extended Data Table~1.

\subsection{Efficiency characterization}
\begin{figure}
\centering
\includegraphics[width=1\linewidth]{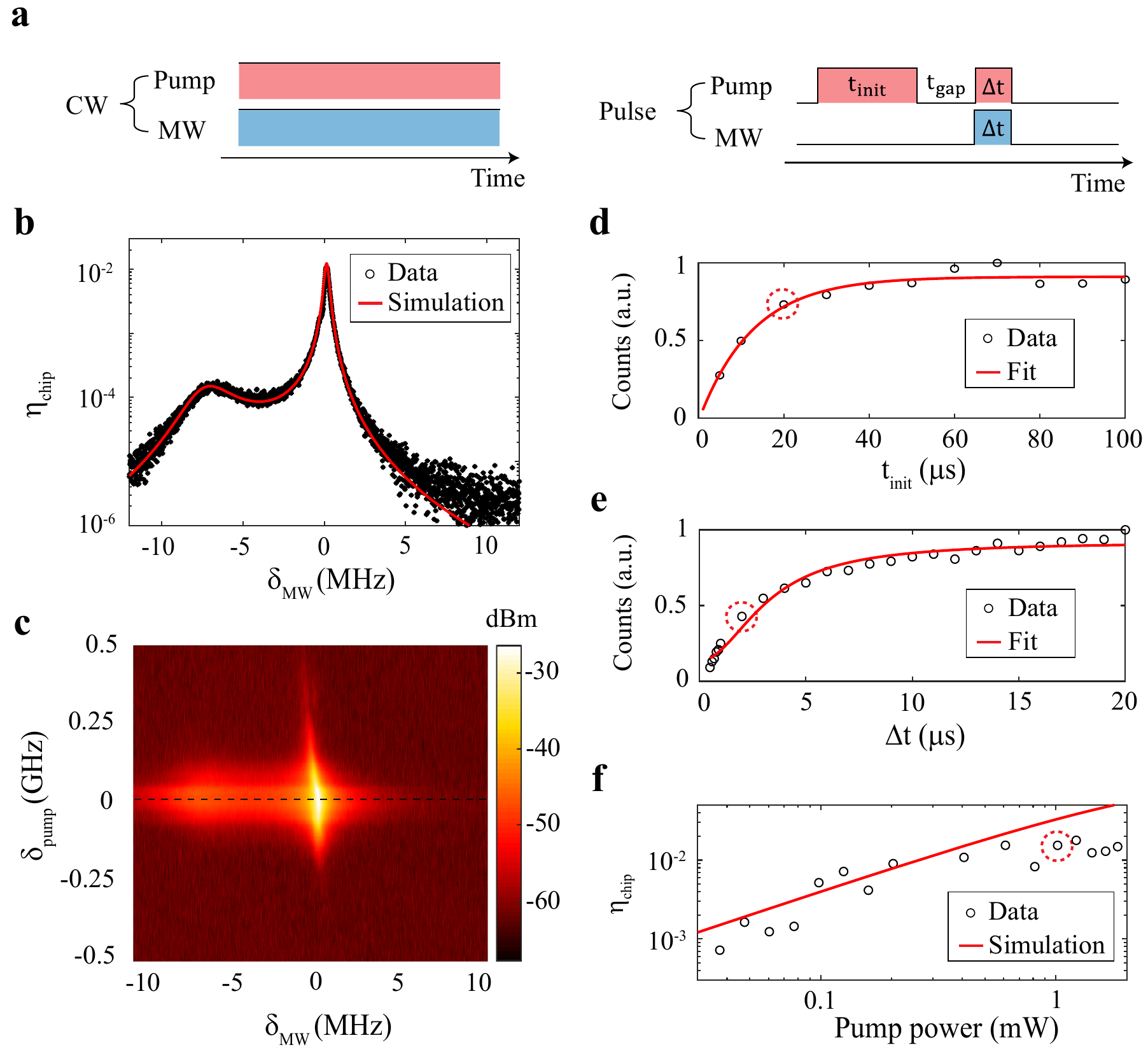}
\caption{\label{Fig2}
Transducer efficiency characterization. (a) CW and pulsed mode operation diagram. (b) CW mode transduction efficiency with $1$~mW optical pump at different input microwave frequencies. The two peaks correspond to the spin at the center and microwave resonator at $\delta_{mw}=2\pi\times -7.1$~MHz (c) Microwave-to-optical transduction signals under various optical pump and microwave input frequencies. The dashed line is the data shown in (b), which is also the pump frequency used in later experiments. (d) Sweep of initialization time $t_{init}$ with transduction time $\Delta t = 2$~$\mu$s, and a phenomenological exponential fit. (e) Bandwidth characterization by sweeping $\Delta t$ with $t_{init}=20$~$\mu$s, and a bandwidth fit in the frequency domain. (f) Pulse mode efficiency with varying pump power, and a simulation. The red circles in (d),(e), and (f) are all under the same experimental conditions, e.g. (e) is measured with $t_{init}=20$~$\mu$s and a pump power of $1$~mW.
}
\end{figure}

\begin{figure*}
\centering
\includegraphics[width=0.95\linewidth]{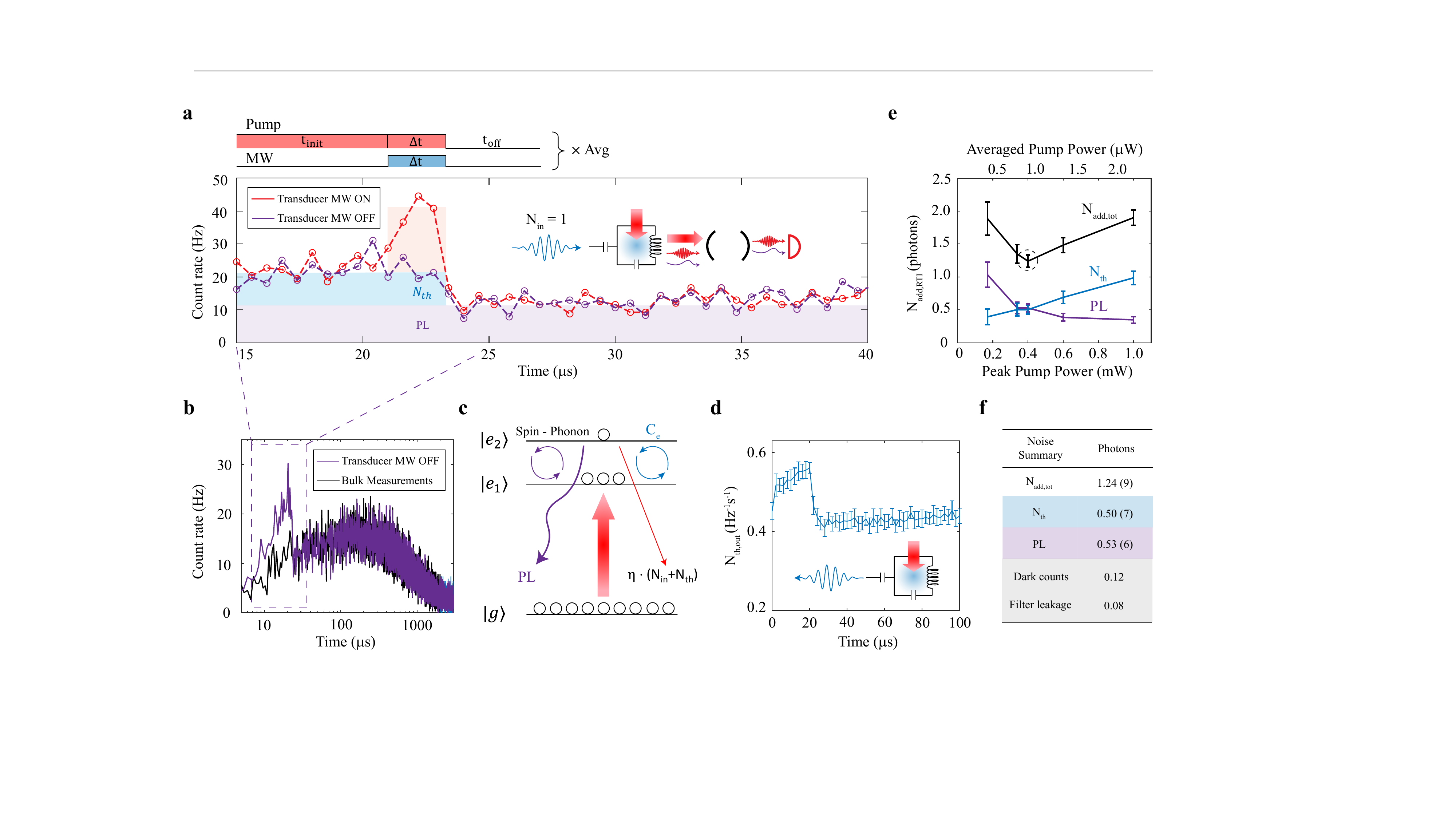}
\caption{\label{Fig3}
Microwave-to-optical transduction noise characterization. (a) Transduced optical photon detection on a superconducting nanowire single photon detector (SNSPD) with a single classical microwave input photon per pulse. The pulse sequence is composed of a $20$~$\mu$s-long initialization time and a $2$~$\mu$s-long transduction probe with $0.4$~mW peak pump power. (b) Zoomed-out histogram of (a), revealing the photo-luminescence (PL) noise in the transduction zone (purple), and PL measured in bulk under the same conditions (black). (c) Diagram of transduction noise composition. An optical pump populates $|e_1\rangle$ and noise photons can be generated in two ways: thermal microwave photons converted out via transduction (blue), and spin-phonon coupling that generates PL (purple). (d) Microwave resonator thermometry with the same transduction pulse sequence. (e) Added noise referred to the input at different pump powers. (f) A summary of different noise source contributions for the data shown in (a).
}
\end{figure*}
Having established coupling between the atomic transitions and their respective fields, we characterize M2O transduction first with a continuous-wave (CW) optical pump and microwave input (Fig.~2a). With optical heterodyne measurements, we obtain a peak chip transduction efficiency of $1.1\%$ under $1$~mW of pump power ($C_a = 0.22$). The narrower peak at the center of Fig.~2b corresponds to the spin transition, and the broader peak at $\delta_{MW} =2\pi\times -7.1$~MHz corresponds to the microwave resonator. The data agrees with the simulation using equation~\ref{eq-eff}. At the same pump power, both input microwave and pump frequencies are scanned to reveal the signal structure, where the maximum appears when both the optical pump and microwave input are on resonance with the atomic transitions (Fig.~2c). Here and in all further experiments, we operate at zero external magnetic fields, where the highest efficiencies are measured (Extended Data Fig.~2).

To minimize the added noise generated from undesirable heating of the system under CW operation, we implement a pulsed protocol. The protocol consists of two parts, optical initialization with duration $t_{init}$ and a transduction probe with duration $\Delta t$ separated from the initialization pulse by $t_{gap}$ (Fig.~2a, right). The initialization pulse is required in order to initialize the system with high enough $C_e$, and to saturate the strong crystal absorption which enables round trip propagation of the pump (Extended Data Fig.~3). To investigate the required initialization time, we scan it with a fixed $\Delta t=2$~$\mu$s, where the increasing trend is due to the combination of the aforementioned optical saturation and increasing population of $|e_1\rangle$ (Fig.~2d). We study the microwave bandwidth of the transducer by sweeping the transduction probe length with fixed $t_{init}=20$~$\mu$s, where a $3$~dB suppression time of $\sim2$~$\mu$s was measured (Fig.~2e), showing a $500$~kHz operation bandwidth. Finally, the pump power is swept, where each data point is optimized for the highest efficiency with respect to the input microwave frequency and initialization time between $10$~$\mu$s and $1$~ms (Fig.~2f). The initialization time must be increased for lower powers in order to sufficiently saturate the atoms and populate the excited state. A simulation (red) is shown on top of the data points. In this power regime the efficiency changes roughly linearly with pump power. To obtain the highest efficiencies, we set $t_{gap}=0$, and the pump and microwave pulses with duration $\Delta t$ are temporally overlapping for all experiments above.

\begin{figure*}
\centering
\includegraphics[width=0.95\linewidth]{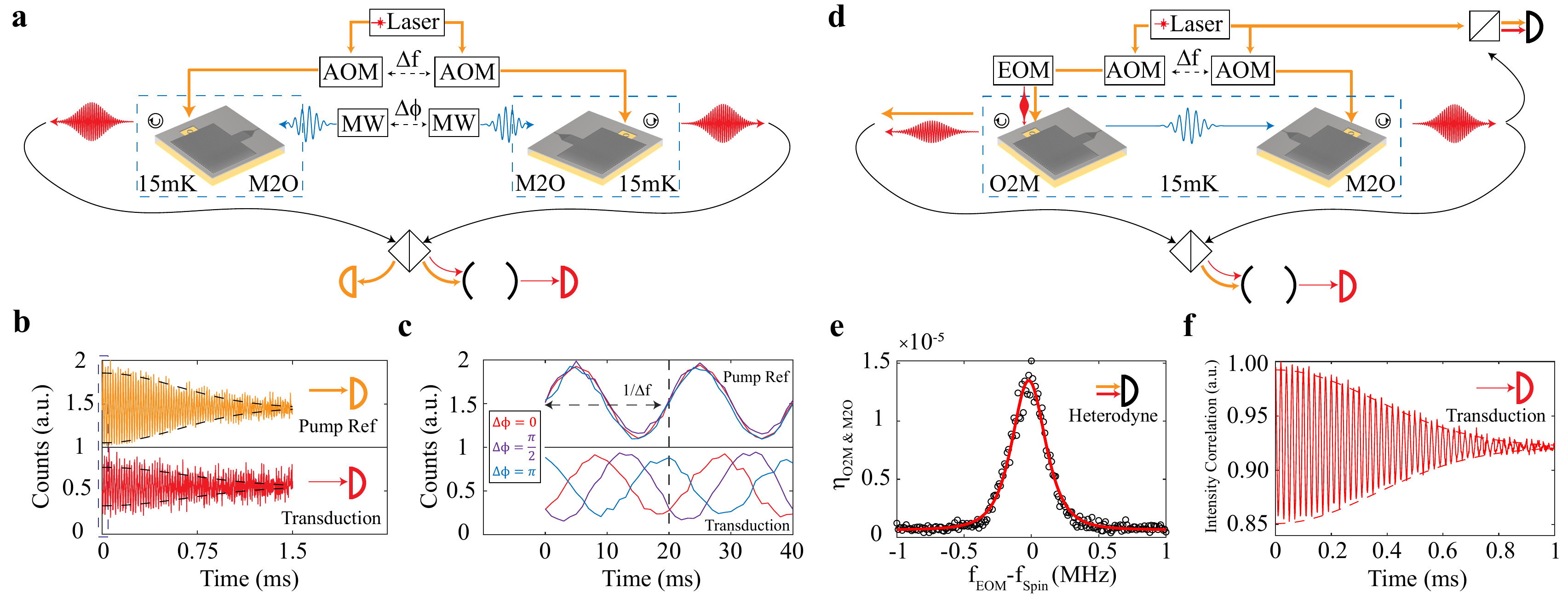}
\caption{\label{Fig4}
Interconnecting transducers. (a) Dual M2O configuration. Two transducers are operated at the same time where the weak input microwave photons are phase shifted by $\Delta \phi$ and optical pumps are frequency shifted by $\Delta f=50$~kHz. The output photons are combined on a beam-splitter and one port is sent to an SNSPD for phase referencing via the pump (b, upper). The other port goes through a narrow frequency filter to spectrally select only the transduction photons, and its interference is detected on another SNSPD (b, lower). A Gaussian decay envelope is observed due to the relative phase drift ($596\pm6$~$\mu$s) between two optical paths. (c) A zoom-in on the interference, where different phase shifts between microwave inputs are implemented and measured. (d) Cascaded O2M to M2O configuration. Two transducers are simultaneously operated where the optical pump is frequency shifted by $\Delta f=50$~kHz. Weak optical photons are sent to the first transducer to perform O2M, from which the second transducer captures the transduced microwave photons and performs M2O. The final optical output is then both measured via heterodyne method and combined with the first input optical light for an intensity correlation measurement. (e) The total O2M-M2O efficiency at different input optical photon frequencies. We use the product of two Lorentzian functions for fitting (red line), which captures two sequential transduction processes. (f) The intensity correlations between the optical input and the transduced output. A Gaussian decay envelope ($359\pm2$~$\mu$s) is observed due to the phase drift from the two separate optical paths.
}
\end{figure*}

However, these parameters can be tuned. By separating the initialization and transduction such that $t_{gap}>0$, we find an exponential decay of efficiency due to the de-population of $|e_1\rangle$ and re-absorption (Extended Data Fig.~4d). More interestingly, as the pump and input photons are first converted to atomic coherence, photons can be temporarily stored in the atoms. We demonstrate this by adding a temporal offset between the optical pump and microwave input during the transduction probe window, and find that the efficiency decays due to the decoherence of the optical and spin transitions, depending on whichever pulse comes first (Extended Data Fig.~4e). This is potentially useful in temporally separating the optical pump, alleviating the need for spectral filtering, and also may lead to a transducer with a built-in quantum memory.

\subsection{Noise characterization}
To characterize the added noise during the transduction process, we send in a single classical microwave photon per pulse and measure the transduced optical photon on a superconducting nanowire single photon detector (SNSPD) through a high-extinction spectral filter that selectively attenuates the reflected pump light. With a $20$~$\mu$s initialization time, $2$~$\mu$s transduction window, and $10$~ms wait time, we measure the transduced photon with an added noise referred to the input of $1.24$~photons ($N_{add, RTI} = 1.24(9)$). Between the zoom-in and zoom-out view of the data as shown in Fig.~3a and b, two distinct noise processes are observed with different time-domain responses, as visualized in Fig.~3c. The population is originally in the ground state where part of it gets excited to $|e_1\rangle$ during the initialization pulse. Even in the absence of an input microwave photon, that population can undergo further processes to produce transduced noise photons.

One type of process is spin-phonon coupling, where the spins couple to a phonon bath, transferring some population to $|e_2\rangle$. Then the population decays from $|e_2\rangle$ to $|g\rangle$, which produces photoluminescence (PL) at the transduction frequency (Supplementary Information). We note that because of the large optical depth, the PL will experience radiation trapping \cite{sumida1994effect} which results in an extended decay time (from $267$ to $672$~$\mu$s) and an increase of the PL before the trapping time. PL measurements in bulk material yield the same count rates (Fig.~3b), thus confirming that this process is independent of the transducer. Furthermore, we calculate an order-of-magnitude estimate of about $1$ to $100$~Hz of PL given an estimated spin-phonon coupling rate and our detection efficiencies (Supplementary Information), which roughly agrees with our measured count rates.

The second type of coupling is spin-resonator coupling, the same mechanism by which input microwave photons are transduced. This noise is only present when the pump is on, in contrast to the PL. To investigate this, we directly measure the microwave output noise power from the chip under the same optical pulse sequence (Fig.~3d). From the calibrated gain and added noise of the output chain, we extract $N_{th, out}$ from resonator thermometry, which roughly matches the detected noise photons in SNSPD measurements (Methods, SI, Extended Data Fig.~5). From this, we conclude that the measured $N_{th}$ in Fig.~3a is transduced noisy photons from the thermal bath of the microwave resonator and waveguide.

To determine the optimal parameters with the smallest added noise summarized in Fig.~3f, we measure the noise at various averaged pump powers. We first sweep the off times $t_{off}$ and observe an increase of the noise at a turning point of around $1$~ms (Extended Data Fig.~6). Returning to $t_{off}=10$~ms, we also sweep the peak pump power (Fig.~3e). In the high-power regime, thermal noise is dominant as expected, whereas in the low-power regime, PL is dominant as efficiency drops quickly due to the de-population and re-absorption processes. Due to this interplay, the optimal peak power is found at an intermediate 400~$\mu$W. Additionally, for applications, high repetition rate is desirable. To increase this while still keeping the noise reasonably low, we can first shorten $t_{off}$ to $2$~ms, then add multiple consecutive transduction pulses after a single initialization. The benefit here is that multiple attempts of transduction can be applied while the effect of initialization is still active, avoiding the noise from additional initialization pulses. We achieve $N_{add,tot}<2$ for $10$ pulses with $2$~ms $t_{off}$, for a repetition rate of $5$~kHz (Extended Data Fig.~7).

\subsection{Interconnecting transducers}
To demonstrate scalability towards large-scale quantum networks, we prepare two transducer chips, where the microwave resonances are offset by $3.4$~MHz from each other due to fabrication disorder. Despite this offset, the most efficient operating point still coincides at the spin frequency (Supplementary Information), showing its resistance to this disorder. We simultaneously do CW M2O transduction on both devices, and interfere the output optical photons (Fig.~4a). We use the reflected pump interference as a reference to correct for the initial phase of the interference of the transduced photons (Fig.~4b, see details in Methods). Here the oscillation frequency is $50$~kHz corresponding to the intentionally added offset between the two pumps, with a Gaussian decay envelope ($\sim596$~$\mu$s) attributed to the phase correlation time between the two independent optical paths.

Then, we tune the phase difference $\Delta\phi$ between the two weak input microwave fields. Since the pump does not sense the phase change, the pump interference overlaps for different $\Delta\phi$ (Fig.~4c, top). On the other hand, $\Delta\phi$ will imprint onto the transduced photons. We measure the phase shift in transduction interference for different $\Delta\phi$, demonstrating the phase-coherent nature of the transducer (Fig.~4c, bottom). This dual M2O-transducer experiment showcases the ease of interconnecting multiple transducers for our REI platform due to its inherently matching optical transitions, and thus its potential for entangling remote quantum nodes.

Furthermore, to demonstrate the ability to optically probe and detect superconducting qubits, we cascade two transducers together with a cryogenic microwave link (Fig.~4d). Optical photons combined with a pump are generated from a weakly driven electro-optical modulator and sent to the first chip to perform O2M transduction. We note that the generated microwave photons can be used to probe qubits, as shown in previous studies \cite{arnold2023all,delaney2022superconducting}. Instead, here we connect them by a microwave link, where the microwave photons are captured by the second transducer and converted back to optical photons, illustrating the ability to optically readout a qubit state. The final transduced optical photons are measured via heterodyne where the total efficiency agrees with the product between O2M and M2O efficiency and chip collection factors (Fig.~4e, Extended Data Fig.~8), where the lack of excess loss shows the advantage of matching spin transitions. By interfering the input optical photons with the final transduced photons (intentionally frequency shifted by $50$~kHz), we measure an intensity correlation between the two independent optical paths, confirming that the frequency was preserved throughout the entire transduction process (Fig.~4f). We note the Gaussian decay envelope of around $359$~$\mu$s, similar to Fig.~4b, which we also attribute to the phase correlation time of the independent optical paths.

\section{Discussion and outlook}
In this work, we present a REI-based on-chip microwave-to-optical transducer. Without an engineered optical cavity, the chip efficiency is $0.76$\% ($3.4\times10^{-5}$ including the pump filtering system) with an added noise down to $1.24(9)$ photons under pulsed operation with $500$~kHz bandwidth. We further demonstrate dual transducer interference experiments that show scalability and the capacity to drive and readout qubits via optical fibers, an essential step towards the remote entanglement of superconducting qubits. The achieved efficiency with low optical and microwave resonator quality factors comes from the ultra-strong effective $\chi^{(2)}$ non-linearity in this REI-based material. Similar efficiency has been achieved for example in LiNbO$_3$, but required optical cavities with quality factors on the order of millions due to the quadratic scaling of efficiency with $\chi^{(2)}$ in the low-cooperativity regime \cite{Fu2021,rueda2016efficient,xu2021bidirectional}. The lack of a high-finesse optical cavity in our device lends itself to scalability in the context of robust fabrication, and also alleviates the need for precise frequency matching. Looking ahead, we identify a few promising paths towards an efficient quantum transducer.

Currently, the added noise is primarily caused by resonator heating from the optical pump and the corresponding PL decay. With a $^{171}$Yb$^{3+}$:YVO$_4$ thin film, a shorter initialization time will be required to optically saturate the substrate, and both thermal and PL noise can be suppressed. Second, the current transducer bandwidth is limited by the narrow spin inhomogeneous linewidth. Working with higher REI densities will not only boost the efficiency, but also increase the bandwidth, as typically the inhomogeneous linewidth increases with dopant density. Increasing the bandwidth will allow us to work with shorter transduction pulses, effectively decreasing the added noise. Meanwhile, PL can be spatially filtered by using a smaller solid angle for collection and also filtered by polarization, as PL is unpolarized.

For efficiency, in addition to doping with higher densities, incorportating an engineered optical cavity is the most direct improvement. The detection efficiency can also be improved by technical means to lower the insertion loss of the pump filtering system. Moving towards hybrid quantum networks, future experiments will be focused on optical access and entanglement generation between remote superconducting qubits. Meanwhile, interfacing superconducting qubits with single REIs or ensembles will also be investigated towards building a hybrid quantum network with disparate physical platforms.

\begin{acknowledgments}
We acknowledge helpful discussions with Keith Schwab, John Bartholomew, Mi Lei, Andrei Ruskuc, Chun-Ju Wu, Sophie Hermans, Adrian Beckert, Tianzhe Zheng, Yiran Gu, Frank Yang, and Srujan Meesala. We thank Matt Shaw and Boris Korzh for help with the SNSPDs, and Yuchun Sun for gold deposition and SEM usage. \textbf{Funding:} This work was primarily supported by Office of Naval Research grant N00014-22-1-2422. We also acknowledge funding from: US Department of Energy, Office of Science, National Quantum Information Science Research Centers, Co-design Center for Quantum Advantage (contract number DE-SC0012704); Gordon and Betty Moore Foundation Experimental Physics Investigators. R. F. acknowledges support from the Quad fellowship. Fabrication was performed in the Kavli Nanoscience Institute at Caltech. \textbf{Author Contributions:} T.X. and R.F. fabricated the devices, performed the measurements and simulations, and analyzed the results. T.X.,R.F., and J.L. built the experimental setup. A.F. supervised the project. T.X., R.F., and A.F. wrote the manuscript, with input from all authors. \textbf{Competing interests:} The authors declare no competing interests. \textbf{Data and materials availability:} The data that support the findings of this study are available in the main text and supplementary materials.
\end{acknowledgments}

\bibliographystyle{naturemag}
\bibliography{ref}

\begin{thebibliography}{10}
\expandafter\ifx\csname url\endcsname\relax
  \def\url#1{\texttt{#1}}\fi
\expandafter\ifx\csname urlprefix\endcsname\relax\def\urlprefix{URL }\fi
\providecommand{\bibinfo}[2]{#2}
\providecommand{\eprint}[2][]{\url{#2}}

\bibitem{kimble2008quantum}
\bibinfo{author}{Kimble, H.~J.}
\newblock \bibinfo{title}{The quantum internet}.
\newblock \emph{\bibinfo{journal}{Nature}} \textbf{\bibinfo{volume}{453}}, \bibinfo{pages}{1023--1030} (\bibinfo{year}{2008}).

\bibitem{cirac1999distributed}
\bibinfo{author}{Cirac, J.~I.}, \bibinfo{author}{Ekert, A.}, \bibinfo{author}{Huelga, S.~F.} \& \bibinfo{author}{Macchiavello, C.}
\newblock \bibinfo{title}{Distributed quantum computation over noisy channels}.
\newblock \emph{\bibinfo{journal}{Physical Review A}} \textbf{\bibinfo{volume}{59}}, \bibinfo{pages}{4249} (\bibinfo{year}{1999}).

\bibitem{arute2019quantum}
\bibinfo{author}{Arute, F.} \emph{et~al.}
\newblock \bibinfo{title}{Quantum supremacy using a programmable superconducting processor}.
\newblock \emph{\bibinfo{journal}{Nature}} \textbf{\bibinfo{volume}{574}}, \bibinfo{pages}{505--510} (\bibinfo{year}{2019}).

\bibitem{lo2014secure}
\bibinfo{author}{Lo, H.-K.}, \bibinfo{author}{Curty, M.} \& \bibinfo{author}{Tamaki, K.}
\newblock \bibinfo{title}{Secure quantum key distribution}.
\newblock \emph{\bibinfo{journal}{Nature Photonics}} \textbf{\bibinfo{volume}{8}}, \bibinfo{pages}{595--604} (\bibinfo{year}{2014}).

\bibitem{pompili2021realization}
\bibinfo{author}{Pompili, M.} \emph{et~al.}
\newblock \bibinfo{title}{Realization of a multinode quantum network of remote solid-state qubits}.
\newblock \emph{\bibinfo{journal}{Science}} \textbf{\bibinfo{volume}{372}}, \bibinfo{pages}{259--264} (\bibinfo{year}{2021}).

\bibitem{baumgratz2016quantum}
\bibinfo{author}{Baumgratz, T.} \& \bibinfo{author}{Datta, A.}
\newblock \bibinfo{title}{Quantum enhanced estimation of a multidimensional field}.
\newblock \emph{\bibinfo{journal}{Physical review letters}} \textbf{\bibinfo{volume}{116}}, \bibinfo{pages}{030801} (\bibinfo{year}{2016}).

\bibitem{Pirandola2018}
\bibinfo{author}{Pirandola, S.}, \bibinfo{author}{Bardhan, B.~R.}, \bibinfo{author}{Gehring, T.}, \bibinfo{author}{Weedbrook, C.} \& \bibinfo{author}{Lloyd, S.}
\newblock \bibinfo{title}{Advances in photonic quantum sensing}.
\newblock \emph{\bibinfo{journal}{Nature Photonics}} \textbf{\bibinfo{volume}{12}}, \bibinfo{pages}{724--733} (\bibinfo{year}{2018}).

\bibitem{google2023suppressing}
\bibinfo{title}{Suppressing quantum errors by scaling a surface code logical qubit}.
\newblock \emph{\bibinfo{journal}{Nature}} \textbf{\bibinfo{volume}{614}}, \bibinfo{pages}{676--681} (\bibinfo{year}{2023}).

\bibitem{kim2023evidence}
\bibinfo{author}{Kim, Y.} \emph{et~al.}
\newblock \bibinfo{title}{Evidence for the utility of quantum computing before fault tolerance}.
\newblock \emph{\bibinfo{journal}{Nature}} \textbf{\bibinfo{volume}{618}}, \bibinfo{pages}{500--505} (\bibinfo{year}{2023}).

\bibitem{lauk2020perspectives}
\bibinfo{author}{Lauk, N.} \emph{et~al.}
\newblock \bibinfo{title}{Perspectives on quantum transduction}.
\newblock \emph{\bibinfo{journal}{Quantum Science and Technology}} \textbf{\bibinfo{volume}{5}}, \bibinfo{pages}{020501} (\bibinfo{year}{2020}).

\bibitem{han2021microwave}
\bibinfo{author}{Han, X.}, \bibinfo{author}{Fu, W.}, \bibinfo{author}{Zou, C.-L.}, \bibinfo{author}{Jiang, L.} \& \bibinfo{author}{Tang, H.~X.}
\newblock \bibinfo{title}{Microwave-optical quantum frequency conversion}.
\newblock \emph{\bibinfo{journal}{Optica}} \textbf{\bibinfo{volume}{8}}, \bibinfo{pages}{1050--1064} (\bibinfo{year}{2021}).

\bibitem{sahu2022quantum}
\bibinfo{author}{Sahu, R.} \emph{et~al.}
\newblock \bibinfo{title}{Quantum-enabled operation of a microwave-optical interface}.
\newblock \emph{\bibinfo{journal}{Nature communications}} \textbf{\bibinfo{volume}{13}}, \bibinfo{pages}{1276} (\bibinfo{year}{2022}).

\bibitem{xu2021bidirectional}
\bibinfo{author}{Xu, Y.} \emph{et~al.}
\newblock \bibinfo{title}{Bidirectional interconversion of microwave and light with thin-film lithium niobate}.
\newblock \emph{\bibinfo{journal}{Nature communications}} \textbf{\bibinfo{volume}{12}}, \bibinfo{pages}{4453} (\bibinfo{year}{2021}).

\bibitem{shen2024photonic}
\bibinfo{author}{Shen, M.} \emph{et~al.}
\newblock \bibinfo{title}{Photonic link from single-flux-quantum circuits to room temperature}.
\newblock \emph{\bibinfo{journal}{Nature Photonics}} \bibinfo{pages}{1--8} (\bibinfo{year}{2024}).

\bibitem{jiang2020efficient}
\bibinfo{author}{Jiang, W.} \emph{et~al.}
\newblock \bibinfo{title}{Efficient bidirectional piezo-optomechanical transduction between microwave and optical frequency}.
\newblock \emph{\bibinfo{journal}{Nature communications}} \textbf{\bibinfo{volume}{11}}, \bibinfo{pages}{1166} (\bibinfo{year}{2020}).

\bibitem{weaver2024integrated}
\bibinfo{author}{Weaver, M.~J.} \emph{et~al.}
\newblock \bibinfo{title}{An integrated microwave-to-optics interface for scalable quantum computing}.
\newblock \emph{\bibinfo{journal}{Nature Nanotechnology}} \textbf{\bibinfo{volume}{19}}, \bibinfo{pages}{166--172} (\bibinfo{year}{2024}).

\bibitem{higginbotham2018harnessing}
\bibinfo{author}{Higginbotham, A.~P.} \emph{et~al.}
\newblock \bibinfo{title}{Harnessing electro-optic correlations in an efficient mechanical converter}.
\newblock \emph{\bibinfo{journal}{Nature Physics}} \textbf{\bibinfo{volume}{14}}, \bibinfo{pages}{1038--1042} (\bibinfo{year}{2018}).

\bibitem{zhao2024quantum}
\bibinfo{author}{Zhao, H.}, \bibinfo{author}{Chen, W.~D.}, \bibinfo{author}{Kejriwal, A.} \& \bibinfo{author}{Mirhosseini, M.}
\newblock \emph{\bibinfo{journal}{arXiv preprint arXiv:2406.02704}}  (\bibinfo{year}{2024}).
\newblock \eprint{2406.02704}.

\bibitem{kumar2023quantum}
\bibinfo{author}{Kumar, A.} \emph{et~al.}
\newblock \bibinfo{title}{Quantum-enabled millimetre wave to optical transduction using neutral atoms}.
\newblock \emph{\bibinfo{journal}{Nature}} \textbf{\bibinfo{volume}{615}}, \bibinfo{pages}{614--619} (\bibinfo{year}{2023}).

\bibitem{rochman2023microwave}
\bibinfo{author}{Rochman, J.}, \bibinfo{author}{Xie, T.}, \bibinfo{author}{Bartholomew, J.~G.}, \bibinfo{author}{Schwab, K.} \& \bibinfo{author}{Faraon, A.}
\newblock \bibinfo{title}{Microwave-to-optical transduction with erbium ions coupled to planar photonic and superconducting resonators}.
\newblock \emph{\bibinfo{journal}{Nature Communications}} \textbf{\bibinfo{volume}{14}}, \bibinfo{pages}{1153} (\bibinfo{year}{2023}).

\bibitem{fernandez2019cavity}
\bibinfo{author}{Fernandez-Gonzalvo, X.}, \bibinfo{author}{Horvath, S.~P.}, \bibinfo{author}{Chen, Y.-H.} \& \bibinfo{author}{Longdell, J.~J.}
\newblock \bibinfo{title}{Cavity-enhanced raman heterodyne spectroscopy in er 3+: Y 2 sio 5 for microwave to optical signal conversion}.
\newblock \emph{\bibinfo{journal}{Physical Review A}} \textbf{\bibinfo{volume}{100}}, \bibinfo{pages}{033807} (\bibinfo{year}{2019}).

\bibitem{mirhosseini2020superconducting}
\bibinfo{author}{Mirhosseini, M.}, \bibinfo{author}{Sipahigil, A.}, \bibinfo{author}{Kalaee, M.} \& \bibinfo{author}{Painter, O.}
\newblock \bibinfo{title}{Superconducting qubit to optical photon transduction}.
\newblock \emph{\bibinfo{journal}{Nature}} \textbf{\bibinfo{volume}{588}}, \bibinfo{pages}{599--603} (\bibinfo{year}{2020}).

\bibitem{delaney2022superconducting}
\bibinfo{author}{Delaney, R.} \emph{et~al.}
\newblock \bibinfo{title}{Superconducting-qubit readout via low-backaction electro-optic transduction}.
\newblock \emph{\bibinfo{journal}{Nature}} \textbf{\bibinfo{volume}{606}}, \bibinfo{pages}{489--493} (\bibinfo{year}{2022}).

\bibitem{arnold2023all}
\bibinfo{author}{Arnold, G.} \emph{et~al.}
\newblock \bibinfo{title}{All-optical single-shot readout of a superconducting qubit}.
\newblock \emph{\bibinfo{journal}{arXiv preprint arXiv:2310.16817}}  (\bibinfo{year}{2023}).

\bibitem{sahu2023entangling}
\bibinfo{author}{Sahu, R.} \emph{et~al.}
\newblock \bibinfo{title}{Entangling microwaves with light}.
\newblock \emph{\bibinfo{journal}{Science}} \textbf{\bibinfo{volume}{380}}, \bibinfo{pages}{718--721} (\bibinfo{year}{2023}).

\bibitem{meesala2024non}
\bibinfo{author}{Meesala, S.} \emph{et~al.}
\newblock \bibinfo{title}{Non-classical microwave--optical photon pair generation with a chip-scale transducer}.
\newblock \emph{\bibinfo{journal}{Nature Physics}} \bibinfo{pages}{1--7} (\bibinfo{year}{2024}).

\bibitem{meesala2023quantum}
\bibinfo{author}{Meesala, S.} \emph{et~al.}
\newblock \bibinfo{title}{Quantum entanglement between optical and microwave photonic qubits}.
\newblock \emph{\bibinfo{journal}{arXiv preprint arXiv:2312.13559}}  (\bibinfo{year}{2023}).

\bibitem{hamze2020design}
\bibinfo{author}{Hamze, A.~K.}, \bibinfo{author}{Reynaud, M.}, \bibinfo{author}{Geler-Kremer, J.} \& \bibinfo{author}{Demkov, A.~A.}
\newblock \bibinfo{title}{Design rules for strong electro-optic materials}.
\newblock \emph{\bibinfo{journal}{NPJ Computational Materials}} \textbf{\bibinfo{volume}{6}}, \bibinfo{pages}{130} (\bibinfo{year}{2020}).

\bibitem{williamson2014magneto}
\bibinfo{author}{Williamson, L.~A.}, \bibinfo{author}{Chen, Y.-H.} \& \bibinfo{author}{Longdell, J.~J.}
\newblock \bibinfo{title}{Magneto-optic modulator with unit quantum efficiency}.
\newblock \emph{\bibinfo{journal}{Physical review letters}} \textbf{\bibinfo{volume}{113}}, \bibinfo{pages}{203601} (\bibinfo{year}{2014}).

\bibitem{zhong2015optically}
\bibinfo{author}{Zhong, M.} \emph{et~al.}
\newblock \bibinfo{title}{Optically addressable nuclear spins in a solid with a six-hour coherence time}.
\newblock \emph{\bibinfo{journal}{Nature}} \textbf{\bibinfo{volume}{517}}, \bibinfo{pages}{177--180} (\bibinfo{year}{2015}).

\bibitem{kindem2018characterization}
\bibinfo{author}{Kindem, J.~M.} \emph{et~al.}
\newblock \bibinfo{title}{Characterization of yb 3+ 171: Yvo 4 for photonic quantum technologies}.
\newblock \emph{\bibinfo{journal}{Physical Review B}} \textbf{\bibinfo{volume}{98}}, \bibinfo{pages}{024404} (\bibinfo{year}{2018}).

\bibitem{bartholomew2020chip}
\bibinfo{author}{Bartholomew, J.~G.} \emph{et~al.}
\newblock \bibinfo{title}{On-chip coherent microwave-to-optical transduction mediated by ytterbium in yvo4}.
\newblock \emph{\bibinfo{journal}{Nature communications}} \textbf{\bibinfo{volume}{11}}, \bibinfo{pages}{3266} (\bibinfo{year}{2020}).

\bibitem{zhou2023photonic}
\bibinfo{author}{Zhou, Z.-Q.} \emph{et~al.}
\newblock \bibinfo{title}{Photonic integrated quantum memory in rare-earth doped solids}.
\newblock \emph{\bibinfo{journal}{Laser \& Photonics Reviews}} \textbf{\bibinfo{volume}{17}}, \bibinfo{pages}{2300257} (\bibinfo{year}{2023}).

\bibitem{hatipoglu2024situ}
\bibinfo{author}{Hatipoglu, U.}, \bibinfo{author}{Sonar, S.}, \bibinfo{author}{Lake, D.~P.}, \bibinfo{author}{Meesala, S.} \& \bibinfo{author}{Painter, O.}
\newblock \bibinfo{title}{In situ tuning of optomechanical crystals with nano-oxidation}.
\newblock \emph{\bibinfo{journal}{Optica}} \textbf{\bibinfo{volume}{11}}, \bibinfo{pages}{371--375} (\bibinfo{year}{2024}).

\bibitem{kindem2020control}
\bibinfo{author}{Kindem, J.~M.} \emph{et~al.}
\newblock \bibinfo{title}{Control and single-shot readout of an ion embedded in a nanophotonic cavity}.
\newblock \emph{\bibinfo{journal}{Nature}} \textbf{\bibinfo{volume}{580}}, \bibinfo{pages}{201--204} (\bibinfo{year}{2020}).

\bibitem{ourari2023indistinguishable}
\bibinfo{author}{Ourari, S.} \emph{et~al.}
\newblock \bibinfo{title}{Indistinguishable telecom band photons from a single er ion in the solid state}.
\newblock \emph{\bibinfo{journal}{Nature}} \textbf{\bibinfo{volume}{620}}, \bibinfo{pages}{977--981} (\bibinfo{year}{2023}).

\bibitem{hedges2010efficient}
\bibinfo{author}{Hedges, M.~P.}, \bibinfo{author}{Longdell, J.~J.}, \bibinfo{author}{Li, Y.} \& \bibinfo{author}{Sellars, M.~J.}
\newblock \bibinfo{title}{Efficient quantum memory for light}.
\newblock \emph{\bibinfo{journal}{Nature}} \textbf{\bibinfo{volume}{465}}, \bibinfo{pages}{1052--1056} (\bibinfo{year}{2010}).

\bibitem{lago2021telecom}
\bibinfo{author}{Lago-Rivera, D.}, \bibinfo{author}{Grandi, S.}, \bibinfo{author}{Rakonjac, J.~V.}, \bibinfo{author}{Seri, A.} \& \bibinfo{author}{de~Riedmatten, H.}
\newblock \bibinfo{title}{Telecom-heralded entanglement between multimode solid-state quantum memories}.
\newblock \emph{\bibinfo{journal}{Nature}} \textbf{\bibinfo{volume}{594}}, \bibinfo{pages}{37--40} (\bibinfo{year}{2021}).

\bibitem{ruskuc2024scalable}
\bibinfo{author}{Ruskuc, A.} \emph{et~al.}
\newblock \bibinfo{title}{Scalable multipartite entanglement of remote rare-earth ion qubits}.
\newblock \emph{\bibinfo{journal}{arXiv preprint arXiv:2402.16224}}  (\bibinfo{year}{2024}).

\bibitem{probst2013anisotropic}
\bibinfo{author}{Probst, S.} \emph{et~al.}
\newblock \bibinfo{title}{Anisotropic rare-earth spin ensemble strongly coupled to a superconducting resonator}.
\newblock \emph{\bibinfo{journal}{Physical Review Letters}} \textbf{\bibinfo{volume}{110}}, \bibinfo{pages}{157001} (\bibinfo{year}{2013}).

\bibitem{kjaergaard2020superconducting}
\bibinfo{author}{Kjaergaard, M.} \emph{et~al.}
\newblock \bibinfo{title}{Superconducting qubits: Current state of play}.
\newblock \emph{\bibinfo{journal}{Annual Review of Condensed Matter Physics}} \textbf{\bibinfo{volume}{11}}, \bibinfo{pages}{369--395} (\bibinfo{year}{2020}).

\bibitem{sumida1994effect}
\bibinfo{author}{Sumida, D.} \& \bibinfo{author}{Fan, T.}
\newblock \bibinfo{title}{Effect of radiation trapping on fluorescence lifetime and emission cross section measurements in solid-state laser media}.
\newblock \emph{\bibinfo{journal}{Optics Letters}} \textbf{\bibinfo{volume}{19}}, \bibinfo{pages}{1343--1345} (\bibinfo{year}{1994}).

\bibitem{Fu2021}
\bibinfo{author}{Fu, W.} \emph{et~al.}
\newblock \bibinfo{title}{Cavity electro-optic circuit for microwave-to-optical conversion in the quantum ground state}.
\newblock \emph{\bibinfo{journal}{Phys. Rev. A}} \textbf{\bibinfo{volume}{103}}, \bibinfo{pages}{053504} (\bibinfo{year}{2021}).

\bibitem{rueda2016efficient}
\bibinfo{author}{Rueda, A.} \emph{et~al.}
\newblock \bibinfo{title}{Efficient microwave to optical photon conversion: an electro-optical realization}.
\newblock \emph{\bibinfo{journal}{Optica}} \textbf{\bibinfo{volume}{3}}, \bibinfo{pages}{597--604} (\bibinfo{year}{2016}).

\end{thebibliography}


\begin{thebibliography}{10}
\expandafter\ifx\csname url\endcsname\relax
  \def\url#1{\texttt{#1}}\fi
\expandafter\ifx\csname urlprefix\endcsname\relax\def\urlprefix{URL }\fi
\providecommand{\bibinfo}[2]{#2}
\providecommand{\eprint}[2][]{\url{#2}}

\bibitem{lei2023many}
\bibinfo{author}{Lei, M.} \emph{et~al.}
\newblock \bibinfo{title}{Many-body cavity quantum electrodynamics with driven inhomogeneous emitters}.
\newblock \emph{\bibinfo{journal}{Nature}} \textbf{\bibinfo{volume}{617}}, \bibinfo{pages}{271--276} (\bibinfo{year}{2023}).

\bibitem{tsang2010cavity}
\bibinfo{author}{Tsang, M.}
\newblock \bibinfo{title}{Cavity quantum electro-optics}.
\newblock \emph{\bibinfo{journal}{Physical Review A}} \textbf{\bibinfo{volume}{81}}, \bibinfo{pages}{063837} (\bibinfo{year}{2010}).

\bibitem{xu2023efficient}
\bibinfo{author}{Xu, H.} \emph{et~al.}
\newblock \bibinfo{title}{Efficient quantum transduction using anti-ferromagnetic topological insulators}.
\newblock \emph{\bibinfo{journal}{arXiv preprint arXiv:2308.09048}}  (\bibinfo{year}{2023}).

\bibitem{steck2007quantum}
\bibinfo{author}{Steck, D.~A.}
\newblock \bibinfo{title}{Quantum and atom optics}  (\bibinfo{year}{2007}).

\bibitem{kindem2018characterization}
\bibinfo{author}{Kindem, J.~M.} \emph{et~al.}
\newblock \bibinfo{title}{Characterization of yb 3+ 171: Yvo 4 for photonic quantum technologies}.
\newblock \emph{\bibinfo{journal}{Physical Review B}} \textbf{\bibinfo{volume}{98}}, \bibinfo{pages}{024404} (\bibinfo{year}{2018}).

\bibitem{xie2021characterization}
\bibinfo{author}{Xie, T.} \emph{et~al.}
\newblock \bibinfo{title}{Characterization of er 3+: Yv o 4 for microwave to optical transduction}.
\newblock \emph{\bibinfo{journal}{Physical Review B}} \textbf{\bibinfo{volume}{104}}, \bibinfo{pages}{054111} (\bibinfo{year}{2021}).

\bibitem{rochman2023microwave}
\bibinfo{author}{Rochman, J.}, \bibinfo{author}{Xie, T.}, \bibinfo{author}{Bartholomew, J.~G.}, \bibinfo{author}{Schwab, K.} \& \bibinfo{author}{Faraon, A.}
\newblock \bibinfo{title}{Microwave-to-optical transduction with erbium ions coupled to planar photonic and superconducting resonators}.
\newblock \emph{\bibinfo{journal}{Nature Communications}} \textbf{\bibinfo{volume}{14}}, \bibinfo{pages}{1153} (\bibinfo{year}{2023}).

\bibitem{hease2020bidirectional}
\bibinfo{author}{Hease, W.} \emph{et~al.}
\newblock \bibinfo{title}{Bidirectional electro-optic wavelength conversion in the quantum ground state}.
\newblock \emph{\bibinfo{journal}{PRX Quantum}} \textbf{\bibinfo{volume}{1}}, \bibinfo{pages}{020315} (\bibinfo{year}{2020}).

\bibitem{Xu2020}
\bibinfo{author}{Xu, M.} \emph{et~al.}
\newblock \bibinfo{title}{Radiative cooling of a superconducting resonator}.
\newblock \emph{\bibinfo{journal}{Physical Review Letters}} \textbf{\bibinfo{volume}{124}}, \bibinfo{pages}{033602} (\bibinfo{year}{2020}).

\bibitem{Abragam2012}
\bibinfo{author}{Abragam, A.} \& \bibinfo{author}{Bleaney, B.}
\newblock \emph{\bibinfo{title}{Electron paramagnetic resonance of transition ions}} (\bibinfo{publisher}{OUP Oxford}, \bibinfo{year}{2012}).

\bibitem{Graf1998}
\bibinfo{author}{Graf, F.~R.}, \bibinfo{author}{Renn, A.}, \bibinfo{author}{Zumofen, G.} \& \bibinfo{author}{Wild, U.~P.}
\newblock \bibinfo{title}{Photon-echo attenuation by dynamical processes in rare-earth-ion-doped crystals}.
\newblock \emph{\bibinfo{journal}{Physical Review B}} \textbf{\bibinfo{volume}{58}}, \bibinfo{pages}{5462} (\bibinfo{year}{1998}).

\end{thebibliography}

\end{document}


\begin{center}
    {\huge Supplementary Materials\par}
\end{center}

\section*{Supplementary Figures}
\begin{figure}[h]
\centering
\includegraphics[width=0.7\linewidth]{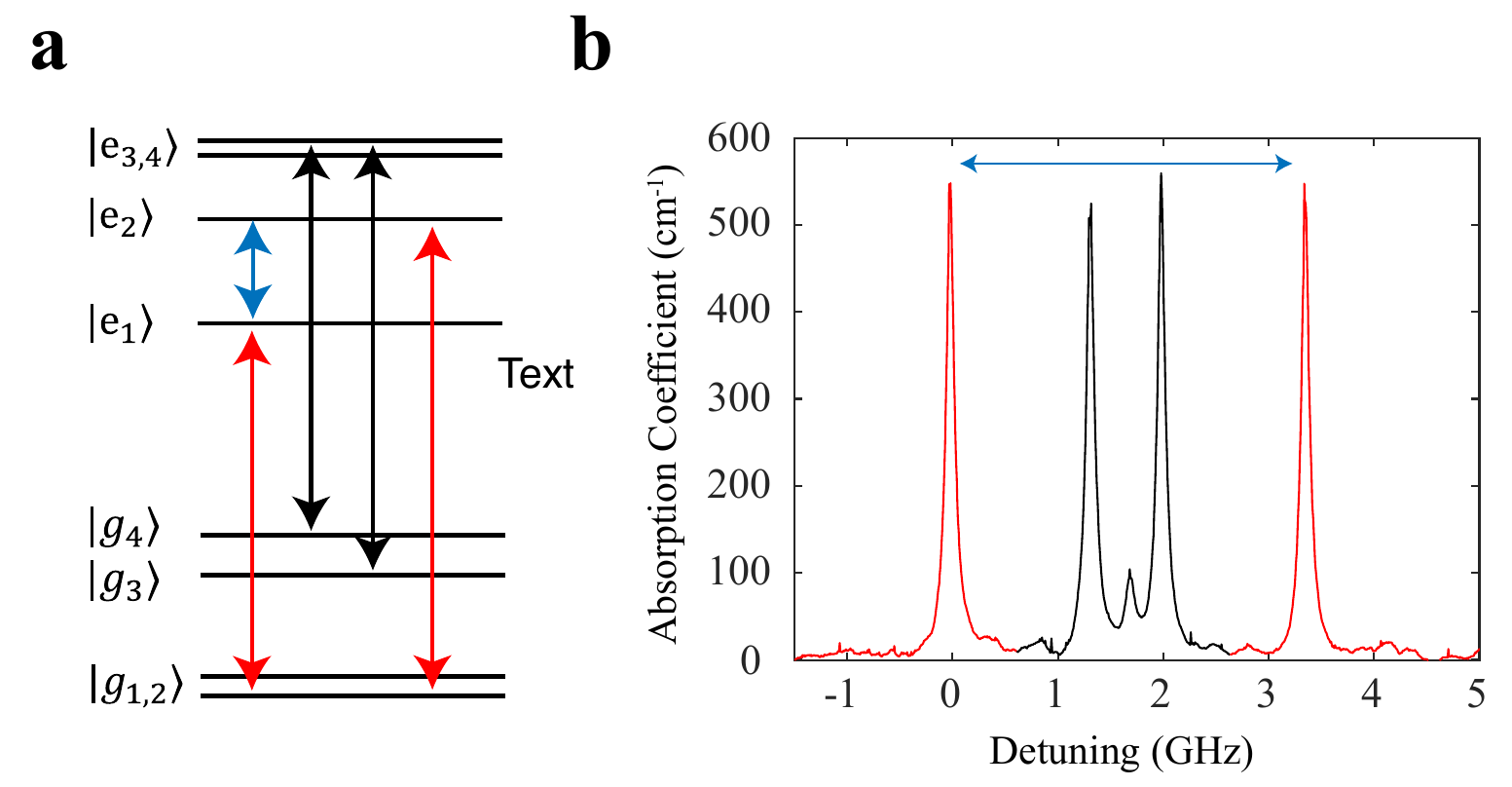}
\caption{\label{ExtFig1}
Full energy levels and optical absorption spectrum. (a) Energy levels of the $^2$F$_{7/2}$(0) ($|g\rangle$) and $^2$F$_{5/2}$(0) ($|e\rangle$) manifolds in $^{171}$Yb$^{3+}$:YVO$_4$. The optical and microwave transitions used in the experiments are labeled red and blue, respectively. (b) Corresponding optical absorption spectrum, where red transitions are used for transduction.
}
\end{figure}

\begin{figure}[h]
\centering
\includegraphics[width=0.7\linewidth]{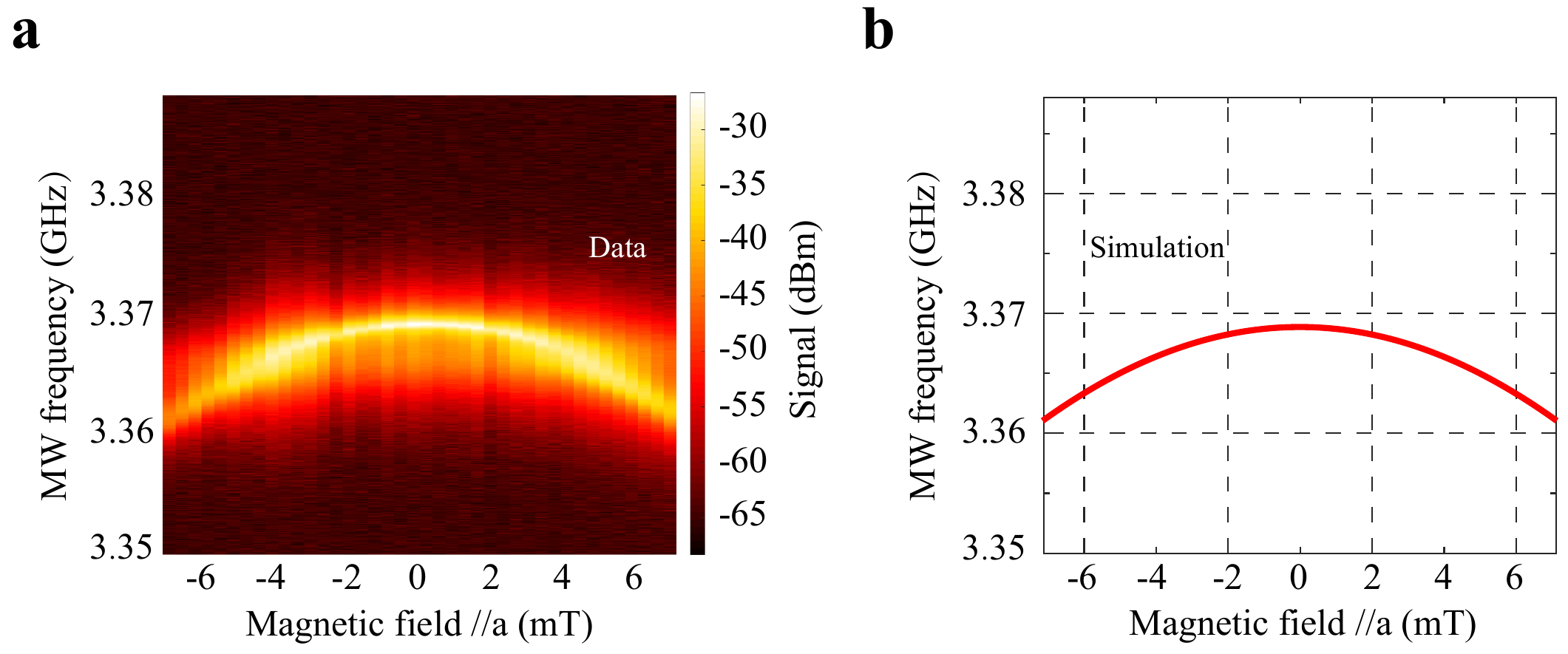}
\caption{\label{ExtFig2}
Microwave to optical transduction signal at different offset magnetic fields. (a) Transduction signal measured by heterodyne with varying offset magnetic field along crystal a-axis. The signal broadens at $\pm$4mT due to the resonant spin and resonator coupling. (b) Simulated excited state spin transition frequency under different offset magnetic fields.
}
\end{figure}

\begin{figure}[h]
\centering
\includegraphics[width=0.9\linewidth]{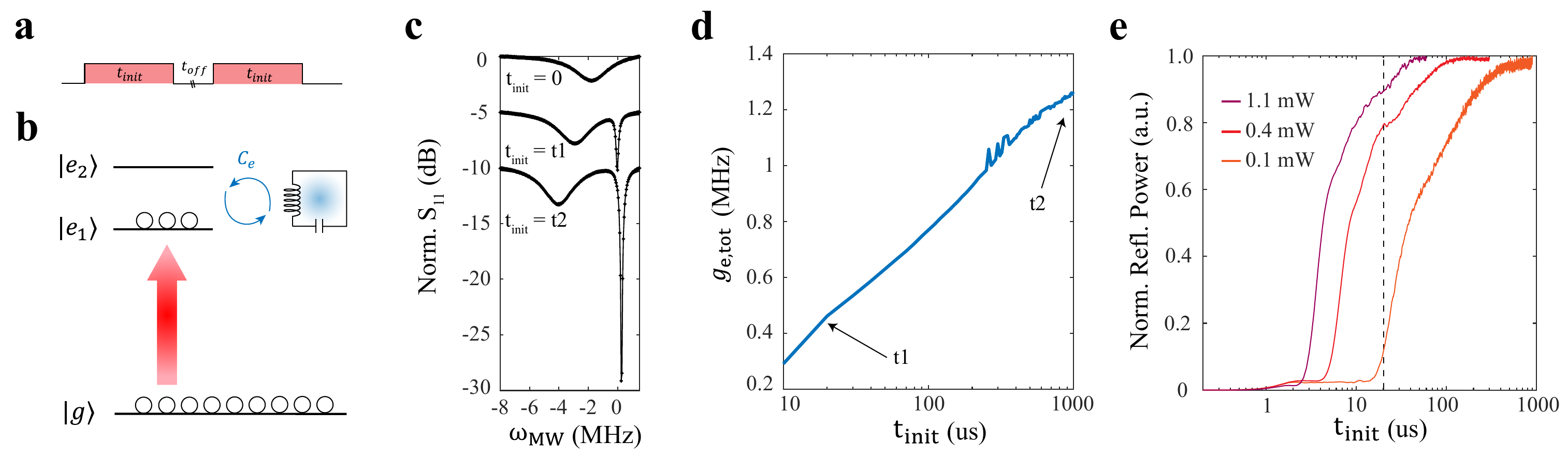}
\caption{\label{ExtFig3}
Optical initialization. (a) Pulse sequence used in the experiments. (b) Diagram of the population transfer process. (c) Microwave reflection spectra at different initialization times. Fits are plotted as solid lines on top of the data. Total coupling strength $g_{e,tot}$ is then extracted from each fit. (d) Fitted total coupling strength $g_{e,tot}$ at various initialization times. (e) Optical saturation effect with different pump powers. The dashed vertical line indicates the initialization time used in the main text (except for Fig.~2f, where initialization time was increased for lower powers).
}
\end{figure}

\begin{figure}
\centering
\includegraphics[width=0.7\linewidth]{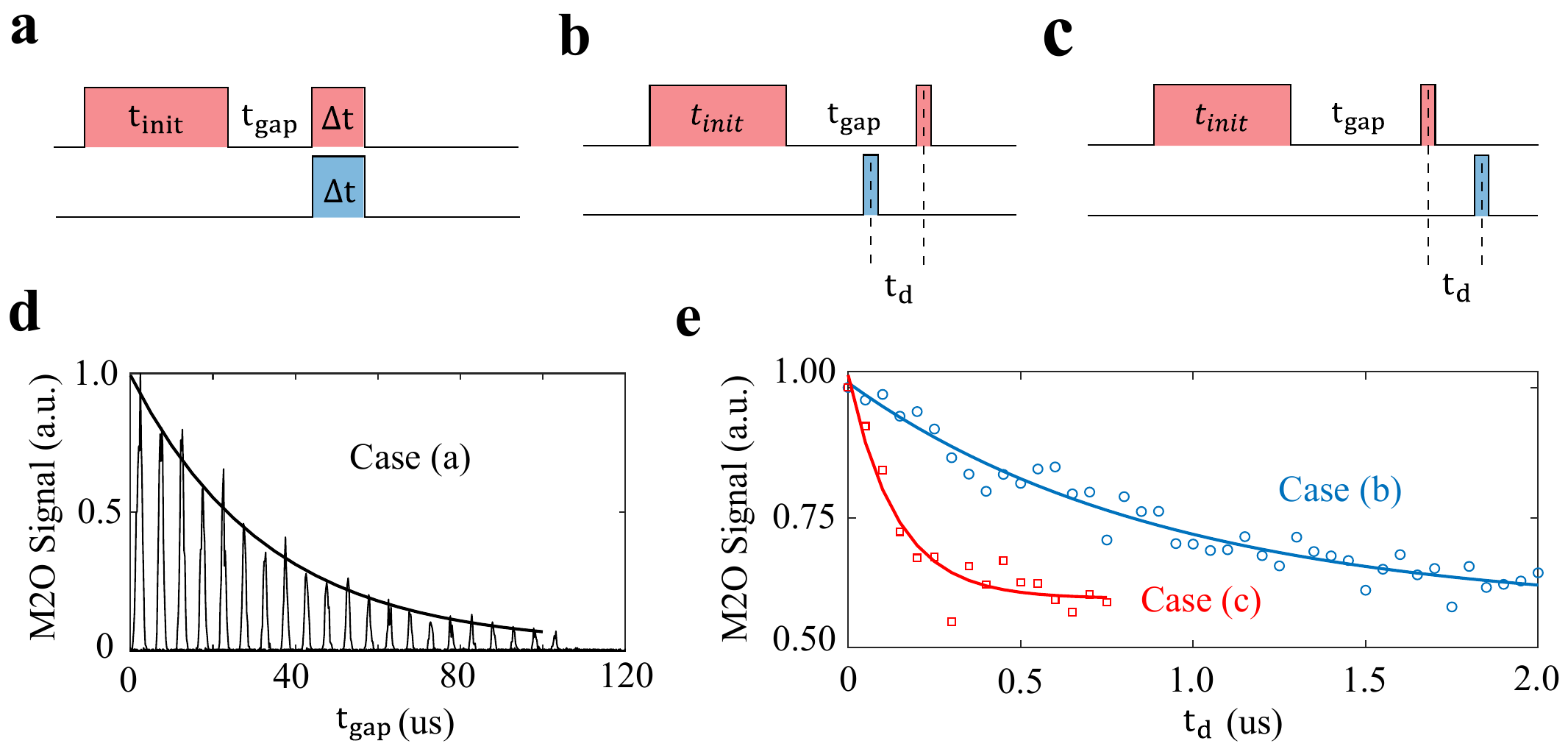}
\caption{\label{ExtFig4}
Different pulse sequences for transduction. (a) A typical pulse sequence used in the main text, but here with $t_{gap} \neq  0$. (b) Transduction pulse sequence with a delay $t_{d}$ between an early microwave input and late optical pump.  (c) Transduction pulse sequence with a delay $t_{d}$ between an early optical pump and a late microwave input. Because the optical/microwave photons will be first converted to optical/spin coherence, a delay to temporally separate the two short transduction pulses will preserve the efficiency as long as it is within the system coherence time.  (d) Experimental data for case (a) with $t_{init} = 20 \mu s$ and $\Delta t = 2 \mu s$. The decreased transduction signal is due to optical T1 decay of the excited state population and also the recovered optical absorption of the pump. (e) Experimental data for cases (b) and (c) with $t_{init} = 20 \mu s$ and $\Delta t = 200 ns$. A $t_{gap}=10\mu s$ is used here to significantly separate the transduction window away from the optical initialization pulse. With an early microwave input, a spin memory time of $T_{m, spin} = 910\pm140$~ns is measured. Whereas with an early optical pump, an optical memory time of $T_{m, op} = 140\pm25$~ns is measured. 
}
\end{figure}

\begin{figure}
\centering
\includegraphics[width=0.9\linewidth]{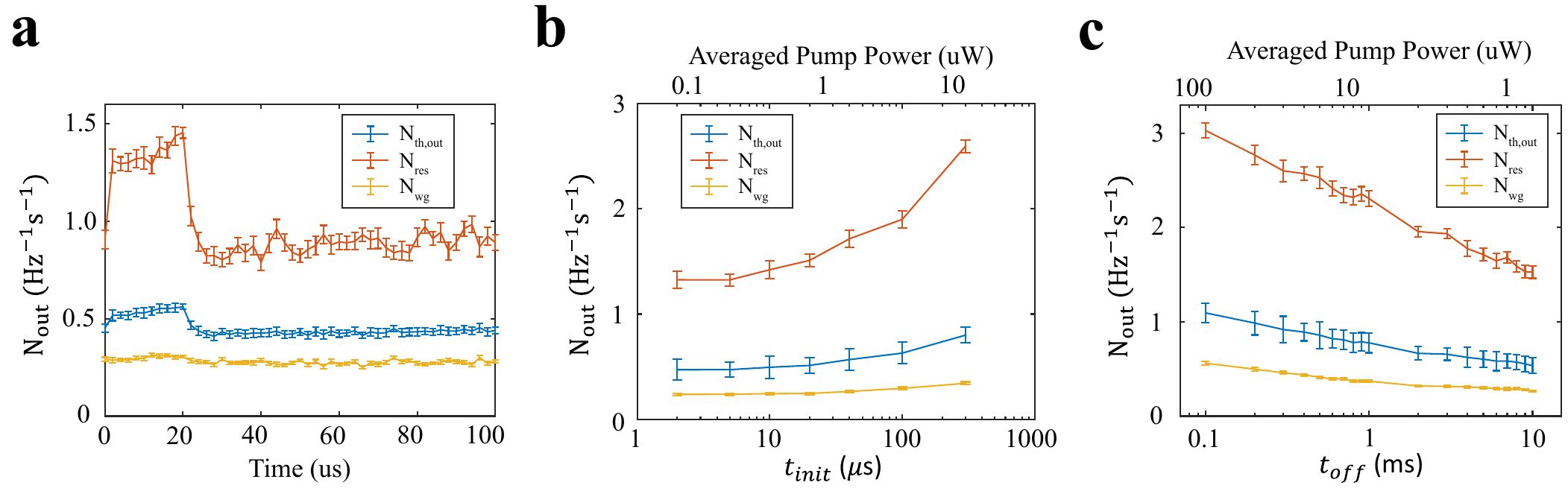}
\caption{\label{ExtFig5}
Detailed transduction noise characterization. (a) $N_{out}$ from microwave thermometry measurements vs. time, where $N_{th,out}$ is the same curve as shown in Fig.~3d. (b) $N_{out}$ at various initialization times with fixed $t_{off}$=10~ms. (c) $N_{out}$ at various $t_{off}$ with fixed $t_{init}$=20~$\mu$s.
}
\end{figure}

\begin{figure}
\centering
\includegraphics[width=0.4\linewidth]{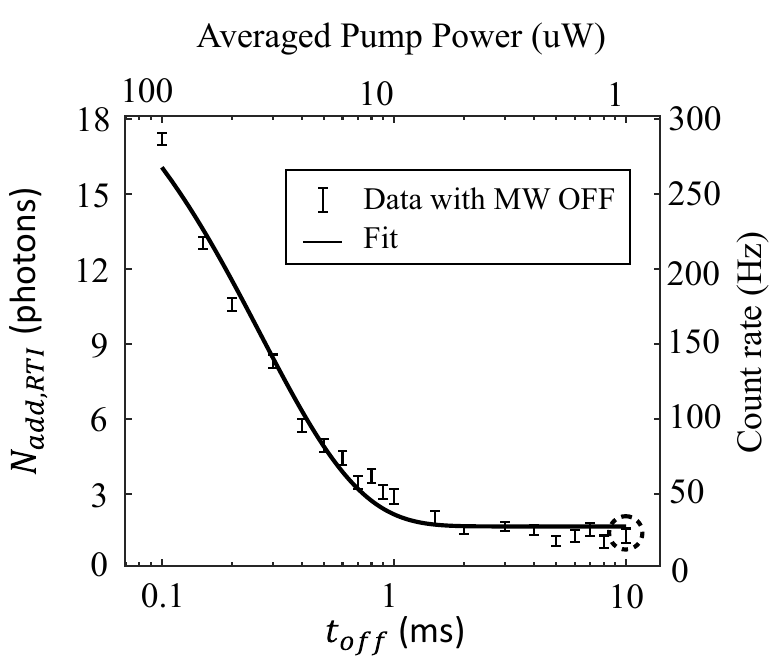}
\caption{\label{ExtFig6}
Total added noise at various $t_{off}$ with fixed $t_{init}$=20~$\mu$s using photon counting measurements. An increase of the noise flux is observed at a turn-over time of around 1~ms. We use a phenomenological exponential curve to fit and capture the key feature.
}
\end{figure}

\begin{figure}[h]
\centering
\includegraphics[width=0.8\linewidth]{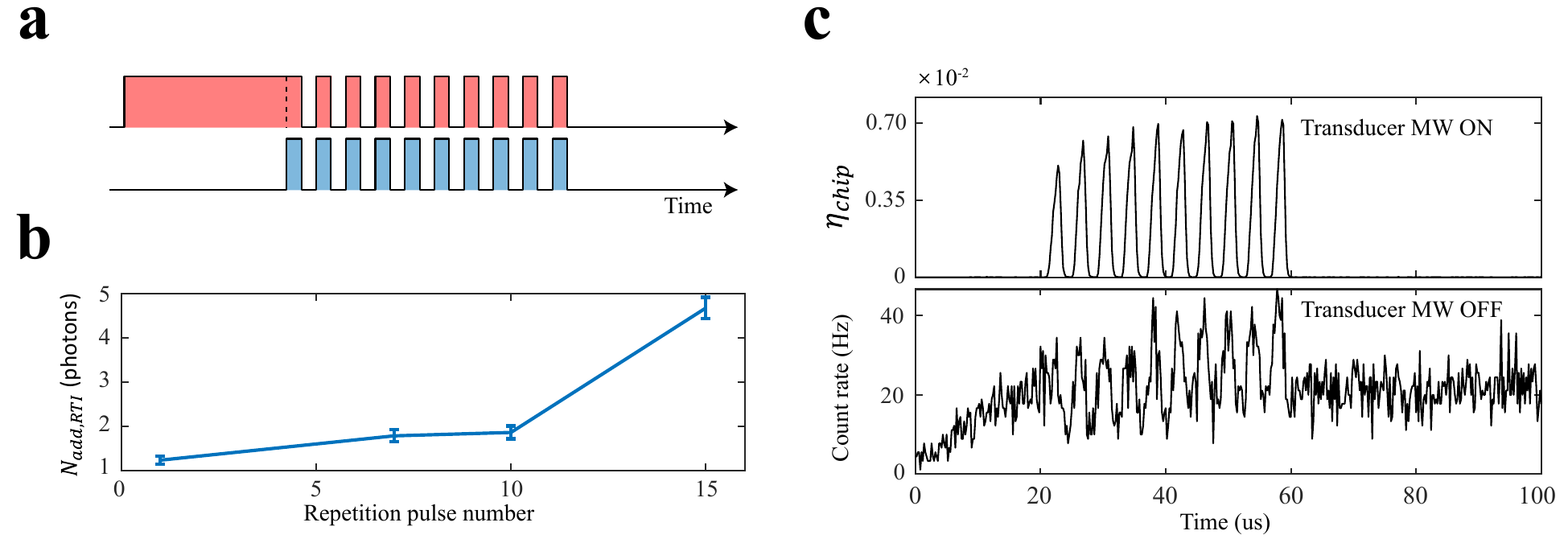}
\caption{\label{ExtFig7}
Consecutive transduction pulses. (a) Pulse diagram of the experiment, consisting of an optical initialization pulse $t_{init}=20$~$\mu$s and multiple transduction probes with $\Delta t = 2$~$\mu$s and a separation of 2~$\mu$s in between. (b) The total added noise for varying pulse repetitions. (c) An example of 10 consecutive transduction pulses. Top panel is with microwave input on for efficiency calibration. Bottom panel is with the microwave input off, measuring the added noise induced by the optical pump.
}
\end{figure}

\begin{figure}[h]
\centering
\includegraphics[width=0.85\linewidth]{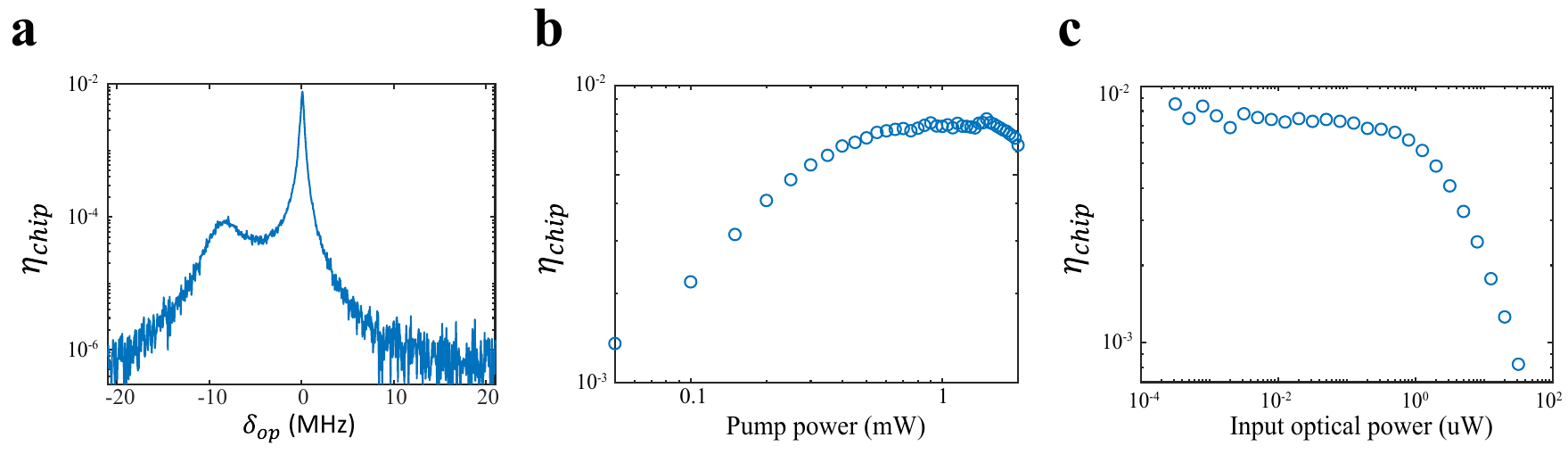}
\caption{\label{ExtFig8}
Optical to microwave transduction. (a) Chip efficiency at various input optical frequencies. The optimal point is when the difference between the input and pump frequencies is equal to the spin frequency. The two peaks corresponds to the spin and resonator contributions, similar to Fig.~2b. Here the pump is on the $|g\rangle$-$|e_1\rangle$ transition, and the input optical photons on the $|g\rangle$-$|e_2\rangle$ transition. (b) Efficiency vs. pump power. (c) Efficiency vs. input optical power. A drop in efficiency is observed when input powers are higher than $\sim 1$~$\mu$W due to the finite number of atoms in the system. 
}
\end{figure}

\begin{figure}
\centering
\includegraphics[width=0.7\linewidth]{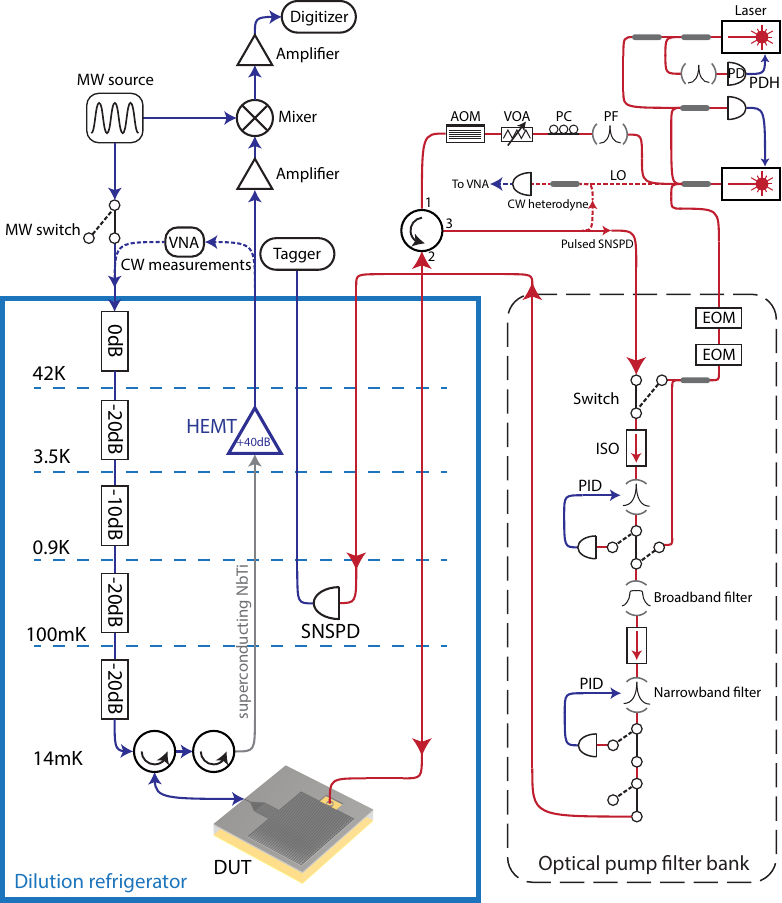}
\caption{\label{ExtFig9}
Experimental setup. BS: beamsplitter of various ratios. PD: photodetector. PF: optical pre-filter. LO: Local oscillator. PC: Polarization controller. PDH: Pound-Drever-Hall locking. VOA: variable optical attenuator. AOM: Acousto-optic modulator. ISO: optical isolator. PID: proportional integral derivative locking. EOM: electro-optic modulator. SNSPD: superconducting nanowire single photon detector. HEMT: high electron-mobility transistor. DUT: device under test. Blue lines indicate microwave cables, red lines indicate optical fibers.
}
\end{figure}

\begin{figure}
\renewcommand{\figurename}{Table}
\setcounter{figure}{0}
\centering
\includegraphics[width=0.8\linewidth]{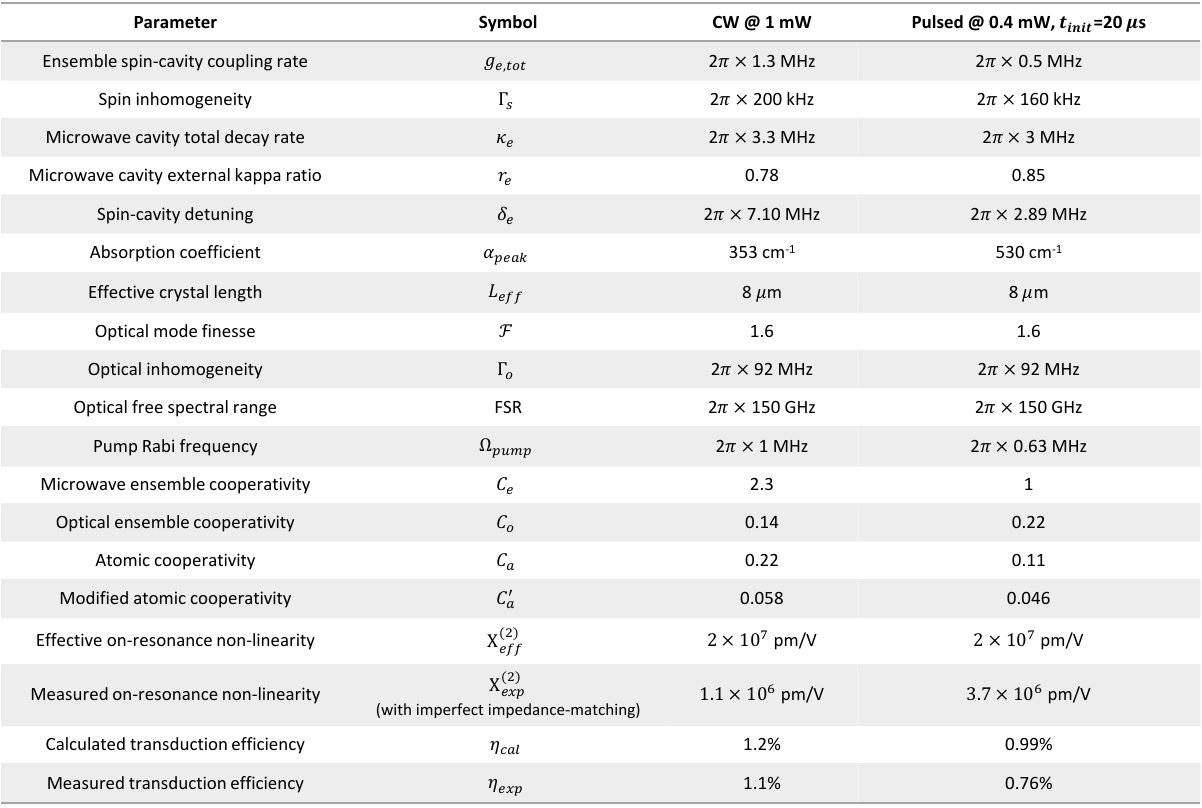}
\caption{\label{ExtTab1}
Relevant parameters of the transducer. $L_{eff}$ is obtained from microwave mode simulation. $\Omega_{pump}$ is calculated from the beam diameter. All other parameters are measured from independent experiments. The difference between the absorption coefficient $\alpha_{peak}$ is because of the difference of total populations inside the transduction manifold between CW and pulse mode. The worse impedance matching in the CW case is because of the large spin-resonator detuning and large ensemble microwave coupling strength. 
}
\end{figure}

\clearpage

\section{Experimental setup}
The experimental setup diagram is shown in Fig.~\ref{ExtFig9}. A master laser (Toptica DL Pro) is locked to a custom reference cavity (Stable laser systems) with $100$~kHz linewidth via Pound-Drever-Hall (PDH) method. The pump laser (Moglabs CEL) is offset phase-locked (with Vescent D2-135) to the master laser. An AOM (Aerodiode model 6) shapes the optical pulses, and additionally provides a small frequency offset, useful for instance in creating the beatnote in Fig.~4a,b. An optical circulator (Precision micro-optics) sends light to the device and collects the reflected light. For CW M2O measurements the reflected light is beat with a local oscillator path from the same pump laser and the beatnote measured on a high speed photodiode (ALPHALAS UPD-35-IR-2), amplified and sent to a vector network analyzer (VNA, Copper Mountain C1209).

For pulsed, low photon count measurements, the reflected light is sent to a pump filtering setup shown in Fig.~\ref{ExtFig9}. Two narrow optical cavities (Stable laser systems) each with a linewidth of $\sim 2$MHz and a free-spectral range of $20$GHz do the majority of the filtering. Theoretically this provides $\sim140$dB of pump extinction, however there is still some residual leakage from unknown sources, possibly defects in fiber components that emit at frequencies outside the bands of the cavities. Thus an additional set of tunable broadband low and high pass filters (Semrock TLP01-995-25X36 and TSP01-995-25X36) placed between the two cavities are used to further extinguish the unwanted photons. Fiber isolators (OF-Link RHPII-980-H6-L-10-FA-Z5) are placed before the cavities to prevent back-reflections. To stabilize the cavities to the transduction frequency, they are periodically offset-locked (1s lock, 5s free-running). Fiber optical switches (Agiltron) are used to switch between the lock and free-running modes, and in the lock mode the cavities are locked to the pump laser, shifted up by the microwave frequency using an EOM (iXBlue NIR-MPX-LN-10). The filtered light is sent to a superconducting nanowire single photon detector (SNSPD, Photonspot), where the events are time-tagged (Swabian Time Tagger 20).

On the microwave side, the microwave input is generated by a CW source (BNC Model 845) and gated by a high-speed switch (Kratos F192A-9) for pulsed operation. The input line in the dilution refrigerator (Bluefors LD250) is attenuated as shown in Fig.~\ref{ExtFig9} for a total of $70$dB attenuation down to the mixing chamber (MXC) plate held at $14$mK with the laser off. Two circulators (LNF-CIC2.8\_4.3A) each with approximately $20$dB isolation block the thermal noise coming from the un-attenuated output line. The second port on the first circulator is connected to the device coplanar waveguide to send in and read out the microwave photons in reflection. The output line consists of superconducting NbTi cables connected from the MXC plate to the HEMT (LNF-LNC2\_6B), which is then connected up to room-temperature amplifiers. See Supplementary Information for detailed calibration on the exact gains and losses of the microwave lines.

For pulsed microwave resonator thermometry measurements, microwave resonator mode occupation is detected via mixing (Marki Microwave IQ0307LXP) the signal coming from the HEMT down to $20$MHz. This is amplified (2x SRS-445A) and measured on a digitizer (Alazar ATS9130).

The device is mounted in the MXC onto a copper post. The light is coupled from a 1060xp fiber to the device and back via a pair of aspheric lenses, focusing the light to a 4$\mu$m beam waist at the gold layer, resulting in a 20$\mu$m beam waist at the chip surface.

For the interconnected transducer experiment (both operating with CW M2O), we simultaneously send in weak microwave photons to each transducer with two synchronized microwave generators. Separate optical pumps are also generated by splitting the pump lasers into two optical paths, each focused onto the circular transduction zone of the respective transducers. Due to the limited laser power, each transducer operates with $\sim400$~$\mu$W pump power. The transduced photons from each transducer, along with the reflected pumps, are collected and combined via a beam-splitter. On one of the output ports we measure the pump photon interference. The other output port goes through a high-extinction spectral filter to another SNSPD to detect only the interference between two transduced photons (Fig.~4a). Based on the phase measured on the pump interference, we correct the initial phase and average the interference over $1000$ data traces for both the pump detection and the transduction detection.

For the cascaded O2M and M2O experiments, separate optical pumps are generated by splitting the pump lasers into two optical paths. The O2M path goes through an EOM and is sent to the O2M chip. Similarly, due to the limited laser power, each transducer operates with $\sim400$~$\mu$W pump power. The output of the final M2O signal is split into two paths for heterodyne detection of the cascaded O2M-M2O, and transduced photon interference through a high-extinction spectral filter.

\section{Total system efficiency}
The total M2O efficiency for the dataset shown in Fig.~3f is $3.4\times10^{-5}$. In addition to the 0.76\% from the chip, this is comprised of the chip to fiber coupling efficiency, pump filtering setup insertion loss, the SNSPD detector efficiency, and various fiber component losses. The chip to fiber coupling efficiency was determined to be approximately 30\%, by measuring the reflected power of an off-resonant CW input. We believe this comes from slight angular misalignment and mode mismatch after reflection, and could be improved with packaging that incorporates tilt control. The circulator port 2 to 3 loss is 70\%. The pump filtering system insertion loss was measured to be 2.5\%. This comes from various individually measured fiber component losses: 2 isolators, each $\sim$80\% transmission; 4 optical switches, each $\sim$80\% transmission; free-space broadband filter, $\sim$50\% transmission due to the fiber coupling; and 2 narrowband fiber-coupled cavities, each $\sim$50\% transmission. The SNSPD detection efficiency is independently measured to exceed 90\%. Finally, by measuring the count rate of a heavily attenuated laser with known power going to the chip and through the filters, we obtain the total detection efficiency of 0.45\%, roughly matching the individually measured component efficiencies.

\section{Device fabrication}
The microwave resonator was fabricated on a 500 $\mu$m thick, 340~ppm doped, c-cut $^{171}$Yb:YVO$_4$ substrate. The doping concentration is measured by secondary ion mass spectrometry. First, 150~nm of niobium was sputtered onto the substrate. Next, the resonator was patterned with MaN-2403 onto the niobium using electron-beam lithography. Then, the pattern was transferred to a 25~nm evaporated aluminum layer after lift-off, which behaves as a hard mask for the subsequent dry etch of niobium by reactive-ion etching in SF$_6$+Ar chemistry. The aluminum was removed with wet etch in aluminum etchant and hydrofluoric acid. Finally, 100~nm of gold was deposited on the back of the chip to form the mirror.

We can also controllably shift the resonance frequency up closer to our spin frequency after measuring once at base temperature. We used a Ga$^+$ focused ion-beam to mill parts of the interdigitated capacitor fingers away, which changed the shunt capacitance and thus the resonance frequency.

\section{Cooperativity definition}
For simplicity, all of the values reported in the main text and Table~S1 refer to the absolute values of the cooperativities at the spin transition where we perform transduction. However, there are detunings in the system which makes the cooperativity a complex number. With the detunings included, we define ensemble microwave cooperativity as $C_e(\delta_e) = \frac{4 g_{e,tot}^2}{\Gamma_e (\kappa_e+2i\delta_e)}$, ensemble optical cooperativity as $C_o(\delta) = \frac{4 g_{o,tot}^2}{\Gamma_o (\kappa_o+2i\delta_o)}$ and atomic cooperativity as $C_a = \frac{4 \Omega_{pump}^2}{\Gamma_o \Gamma_e}$. Here, $g_{e/o,tot}$ is the ensemble microwave/optical coupling strength, $\delta_{e/o}$ is the atom-resonator detuning in microwave or optical domain, and $\Omega_{pump}$ is the single-ion optical Rabi frequency from the pump light.

\section{Calibration of input microwave photons}
We use several different methods to calibrate the total loss of the input microwave chain in order to accurately obtain the M2O efficiency. First, all room-temperature components can be independently measured. Whenever possible total gain/loss of multiple components together is measured to take into account the connector loss, although this is typically smaller than 0.1 dB. The challenge is accurately measuring cryogenic components at base temperature, which we outline below. We note that use two different cryogenic low-noise amplifiers, LNF-LNC-2-6B (HEMT1) and LNF-LNC03-14A (HEMT2), both in the same microwave lines in the dilution fridge but in separate cooldowns. Uncertainties in directly measured quantities below are the fluctuations in the measurements apparatus (i.e. the spectrum analyzer reading), typically 0.1dB. Uncertainties in the HEMT thermometry calibration are obtained from the fit.

First, we connect two identical cryogenic microwave (MW) lines with cryogenic 0 dB attenuators together at the MXC with a short cable, forming a loop. We measure S21 through this loop of -12.3(1)dB, giving an individual line loss of 6.15(5)dB. We then add HEMT2 to one of these lines, where we measure S21 of +27.03(1)dB. This gives a HEMT2 gain of $27+12.3=39.3(1)$ dB. Independently, we also calibrate the gain of HEMT2 using the noise thermometry method outlined in section 3.1. We obtain a total gain of 33.1(2)dB, and with the previously measured cable loss of 6.15(5)dB, give the HEMT2 gain as 39.3(2)dB, matching the other calibration method. This gives confidence in the HEMT thermometry calibration method, as well as the cable loss number.

The above HEMT2 thermometry calibration was done with the 70dB attenuators on the input line with the chip and circulators all in the line. Hence we measure the chip S21 of -43.3(1)dB. This is extrapolated by the S21 at 3.3684 GHz in the absence of the resonator by doing a linear fit with the points of resonance. Subtracting the HEMT gain and adding back the output line cable loss, we obtain the input line loss as $43.3+39.3-6.15=76.5(2)$dB.

Next, we use HEMT1 and repeat the same exercise as above. The S21 through the chip is measured to be -41.07(5)dB, and HEMT1 thermometry calibration gives a total gain of 35.5(1)dB. This gives the input line loss as $41.07+35.5=76.6(2)$dB, matching well with the HEMT1 measurements. For pulsed thermometry we mix down and digitize the output power, for which we do both noise and S21 measurements, giving us another method of calibration. For this, we get S21=+7.1dB(1), and an output line total gain of 86.25(3)dB, giving a total input line loss of $86.25-7.1=79.2(2)$dB. This does not include a MW switch, DC block, and some extra cables, which we measure the total loss to be 2.9dB. This gives the final input line loss as $79.2-2.9=76.3(2)$dB, agreeing with the two previous measurements.

For M2O experiments we additionally use a calibrated 62dB attenuator chain and the MW switch, giving a final value of 141.4(3)dB input line attenuation, from a calibrated MW source outputting a known power. We note that we estimate negligible loss from the chip wirebonds, as the chip S21 agrees with independently measured cable losses without excess loss.

\section{Transduction theory}
\subsection{Setup of the problem}
The system Hamiltonian with both microwave and optical cavity coupling is:
\begin{equation}
\begin{split}
{\cal H} = & \hbar \omega_{o,c} a^{\dagger}a + \hbar \omega_{e,c} b^{\dagger}b + \sum_{k} E_{g,k}\sigma_{gg,k} + \sum_{k} E_{e_1,k}\sigma_{e_1e_1,k} + \sum_{k} E_{e_2,k}\sigma_{e_2e_2,k}
\\
& + \sum_{k} \hbar g_{o,k} (a^{\dagger} \sigma_{o,k}^-+ h.c.) + \sum_{k} \hbar g_{e,k} (b^{\dagger} \sigma_{s,k}^- + h.c.) + \sum_{k} \hbar \Omega_{k} (\sigma_{p,k}^-+h.c.),
\end{split}
\end{equation}
where $a$ is the optical mode at frequency $\omega_{o,c}$, $b$ is the microwave mode at frequency $\omega_{e,c}$, $E$ are the energies of the level denoted in the subscript, $g$ are the optical or microwave coupling rates, $\Omega$ is the pump Rabi frequency, and $\sigma$ are the atomic operators for the transition denoted in the subscript (e.g. $\sigma_{p,k}^-=\ket{g}\bra{e_1}$ for the k-th atom). Implementing the Heisenberg equations of motion, we get
\begin{equation}
\begin{split}
\dot{a} = -\frac{i}{\hbar}[a,{\cal H}] & = -i\omega_{o,c}a - \frac{\kappa_o}{2} a -i \sum_{k}g_{o,k} \sigma_{o,k}^- + \sqrt{\kappa_{o,ext}}a_{in,ext} + \sqrt{\kappa_{o,int}}a_{in,int}
\\
\dot{\sigma}_{o,k}^- = -\frac{i}{\hbar}[\sigma_{o,k}^-,{\cal H}] & = -i\omega_{o,k}\sigma_{o,k}^- -\frac{\gamma_{o}}{2}\sigma_{o,k}^- - ig_{o,k}(\sigma_{gg,k}-\sigma_{e_2 e_2,k})a+i\Omega_{k}\sigma_{s,k}^- -ig_{e,k}\sigma_{p,k}^-b
\\
\dot{\sigma}_{s,k}^- = -\frac{i}{\hbar}[\sigma_{s,k}^-,{\cal H}] & = -i\omega_{s,k}\sigma_{s,k}^- -\frac{\gamma_{s}}{2}\sigma_{s,k}^- - ig_{e,k}(\sigma_{e_1 e_1,k}-\sigma_{e_2 e_2,k})b+i\Omega_{k}\sigma_{o,k}^- -ig_{o,k}\sigma_{p,k}^+ a
\\
\dot{b} = -\frac{i}{\hbar}[b,{\cal H}] & = -i\omega_{e,c}b - \frac{\kappa_e}{2} b -i \sum_{k}g_{e,k} \sigma_{s,k}^- +
\sqrt{\kappa_{e,ext}}b_{in,ext} + \sqrt{\kappa_{e,int}}b_{in,int},
\end{split}
\end{equation}
where we insert $\gamma$ and $\kappa$, the atomic dephasing rates and cavity decay rates respectively. Here, we make two assumptions for simplicity: (1) We assume $\left< a \sigma_{ii}\right> = \left< a\right>\left< \sigma_{ii}\right>$, where quantum correlations have been ignored \cite{lei2023many}. (2) We ignore the last terms for $\dot{\sigma}_{(s,o),k}^-$, as they simply impose a frequency shift that does not affect the conversion process (e.g. $a\sim\sigma_{o}^-,\sigma_{p}^+ a \sim \sigma_{s}^-$). 

With the above assumptions, we can rewrite the differential equations into a matrix with the following definitions:
\begin{equation}
\begin{split}
V_m &= (a,\sigma_{o,1}^-,\cdot\cdot\cdot,\sigma_{o,k}^-,\sigma_{s,1}^-,\cdot\cdot\cdot,\sigma_{s,k}^-,b)^T
\\
V_{m,in} &= (a_{in,ext},a_{in,int},0,\cdot\cdot\cdot,0,0,\cdot\cdot\cdot,0,b_{in,ext},b_{in,int})^T
\\
V_{m,out} &= (a_{out,ext},a_{out,int},0,\cdot\cdot\cdot,0,0,\cdot\cdot\cdot,0,b_{out,ext},b_{out,int})^T
\end{split}
\end{equation}
where $V_m$ is a vector with 2k+2 elements, $V_{m,in/out}$ is a vector with 2k+4 elements. Next we rewrite the system dynamics in the frequency domain, go into the rotating frame of the input optical and microwave frequencies, and use the input-output formalism. The equations of motion will become:
\begin{equation}
\begin{split}
\dot{V}_m &= A V_m + B V_{m,in}
\\
V_{m,out} &= B^T V_m-V_{m,in}
\\
V_{m,out} &= S V_{m,in}
\\
S &= B^T[-i\omega I_{2k+2}A]^{-1}B-I_{2k+4},
\end{split}
\end{equation}
where $I_{d}$ is the d-dimensional identity matrix. The system can then be solved via matrix operations, and the conversion efficiency is an element of the total scattering matrix S (eg: $\eta_{M2O} = |S_{1,5}|^2$).

\subsection{Single atom case}
To gain some physical intuition, we start by just considering a single atom:
\begin{equation}
\begin{split}
A &= 
\begin{pmatrix}
-i\delta_{oc}-\frac{\kappa_o}{2} & -ig_o & 0 & 0 \\
-ig_o(n_{g}-n_{e_2}) &  -i\delta_o-\frac{\gamma_{o}}{2} & i\Omega & 0 \\
0 & i\Omega & -i\delta_e-\frac{\gamma_{s}}{2} &-ig_e(n_{e_1}-n_{e_2})  \\
0 & 0 & -ig_e & -i\delta_{e,c}-\frac{\kappa_e}{2} \\
\end{pmatrix}
\\
B &=
\begin{pmatrix}
\sqrt{\kappa_{o,ext}} & \sqrt{\kappa_{o,int}} & 0 &0  &0  & 0 \\
0 &0  &0  &0  &0  &0  \\
0 &0  &0  &0  &0  &0  \\
0 & 0 & 0 & 0 & \sqrt{\kappa_{e,ext}} & \sqrt{\kappa_{e,int}} \\
\end{pmatrix}
\end{split}
\end{equation}
Setting the detunings to zero, we obtain the microwave to optical conversion efficiency as:
\begin{equation}
\begin{split}
\eta_{M2O} = |S_{1,5}|^2 &= \frac{\kappa_{o,ext}}{\kappa_o}\frac{\kappa_{e,ext}}{\kappa_e} \frac{n_{e1}-n_{e2}}{n_{g}-n_{e2}}\frac{4C_{e,s}C_{o,s}C_{a,s}}{[(1+C_{e,s})(1+C_{o,s})+C_{a,s}]^2}
\\
&= \frac{\kappa_{o,ext}}{\kappa_o}\frac{\kappa_{e,ext}}{\kappa_e} \frac{n_{e1}-n_{e2}}{n_{g}-n_{e2}}\frac{C_{e,s}}{1+C_{e,s}} \frac{4*C_{a,s}^{'}}{(1+C_{a,s}^{'})^2} \frac{C_{o,s}}{1+C_{o,s}},
\end{split}
\end{equation}
where the cooperativity is the single-atom cooperativity with dephasing rate $\gamma$ in the denominators ($C_{o,s} = \frac{4g_o^2(n_g-n_{e_2})}{\kappa_o\gamma_o}$, $C_{e,s} = \frac{4g_e^2(n_{e_1}-n_{e_2})}{\kappa_e\gamma_e}$, $C_{a,s} = \frac{4\Omega^2}{\gamma_o\gamma_e}$). $C_{a,s}^{'}$ is again the modified atomic cooperativity with $(1+C_{e,s})(1+C_{o,s})$ in the denominator. 

The first two terms are the external coupling rate of each resonator. The third term appears due to the fermionic nature of the atoms, and depends on the population distribution between the three energy levels. However, if the optical pumping is strong enough such that it equalizes the pump transition ($n_g = n_{e_1}=\frac{1}{2}$, $n_{e_2}=0$), the third term goes to unity. This makes sense since in this scenario, the population of the two fermionic modes (optical and spin) are equivalent and their effects cancel each other out. The remaining three terms involving the cooperativities are the efficiencies which can be understood as the three stages as shown in the main text in Fig.~1a: a microwave photon to spin coherence, spin coherence to optical coherence (same as electro-optics), and optical coherence to an optical photon.

Also, since the microwave coplanar waveguide connects to the resonator with rate $\kappa_{e,ext}$ and the resonator internal reservoir contributes with rate $\kappa_{e,int}$, the contribution to the microwave noise from the resonator thermal reservoir is:
\begin{equation}
\begin{split}
N_{measured} &= |S(1,5)|^2 N_{wg} +  |S(1,6)|^2 N_{res,int}
\\
N_{add,RTI} &= \frac{N_{measured}}{\eta} = N_{wg} + \frac{|S(1,6)|^2}{|S(1,5)|^2} N_{res,int}
\\
&= N_{wg} + \frac{\kappa_{e,int}}{\kappa_{e,ext}} N_{res,int}
\end{split}
\end{equation}

\begin{figure*}
\setcounter{figure}{9}
\centering
\includegraphics[width=0.9\linewidth]{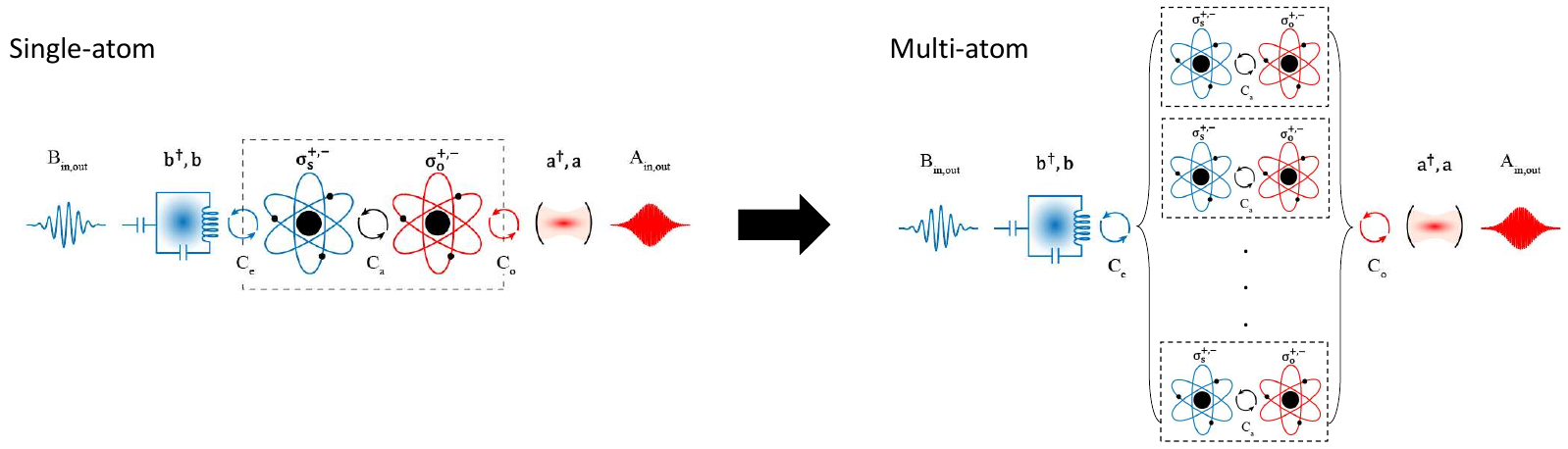}
\caption{\label{S10}
Conversion diagram for single-atom and multi-atom case. In the multi-atom case, each atom can individually couple to the resonator mode, but the mode only couples to the entire ensemble.
}
\end{figure*}

\subsection{Multiple atoms case}
Now we consider the more general ensemble case, where there are multiple atoms that can each individually couple to the optical and microwave resonator mode. However, as we have ignored quantum correlations and consider only the atoms as a collective, the resonator couples to the total contribution of the ensemble, illustrated in Fig.~\ref{S10}. In this case, the matrix can be rewritten as:
\begin{equation}
\begin{split}
A &= 
\begin{pmatrix}
-i\delta_{oc}-\frac{\kappa_o}{2} & -ig_{o,1}  & \cdot \cdot \cdot & -ig_{o,k} &  &  &  &  \\
-ig_{o,1}(n_{g,1}-n_{e_2,1}) & -i\delta_{o,1}-\frac{\gamma_{o}}{2} &  &  & i\Omega_1 &  &  &  \\
\cdot \cdot \cdot &  & \cdot \cdot \cdot &  &  & \cdot \cdot \cdot &  &  \\
-ig_{o,k}(n_{g,k}-n_{e_2,k}) &  &  & -i\delta_{o,k}-\frac{\gamma_{o}}{2} &  &  &i\Omega_k  &  \\
 & i\Omega_1  &  &  & -i\delta_{e,1}-\frac{\gamma_{e}}{2} &  &  & -ig_{e,1}(n_{e_1,1}-n_{e_2,1}) \\
 &  & \cdot \cdot \cdot &  &  & \cdot \cdot \cdot &  & \cdot \cdot \cdot \\
 &  &  & i\Omega_k  &  &  & -i\delta_{e,k}-\frac{\gamma_{e}}{2} & -ig_{e,k}(n_{e_1,k}-n_{e_2,k}) \\
 &  &  &  & -ig_{e,1} & \cdot \cdot \cdot & -ig_{e,k} & -i\delta_{ec}-\frac{\kappa_e}{2} \\
\end{pmatrix}
\\
B &= 
\begin{pmatrix}
\sqrt{\kappa_{o,ext}} & \sqrt{\kappa_{o,int}}  &  &  &  &  &  &  &  &  \\
 &  & 0 &  &  &  &  &  &  &  \\
 &  &  & \cdot \cdot \cdot &  &  &  &  &  &  \\
 &  &  &  & 0 &  &  &  &  &  \\
 &  &  &  &  & 0 &  &  &  &  \\
 &  &  &  &  &  & \cdot \cdot \cdot &  &  &  \\
  &  &  &  &  &  &  & 0 &  &  \\
 &  &  &  &  &  &  &  &  \sqrt{\kappa_{e,ext}} & \sqrt{\kappa_{e,int}}  \\
\end{pmatrix}
\end{split}
\end{equation}
where A is a $(2k+2) \times (2k+2)$ matrix and B is a $(2k+2) \times (2k+4)$ matrix. 

To get a sense of the analytical form of the conversion efficiency, we can consider just two identical atoms. Then, A will be a $6 \times 6$ matrix and B a $6 \times 8$ matrix. Assuming no detunings again, the matrix will become:

\begin{equation}
\begin{split}
A &= 
\begin{pmatrix}
-\frac{\kappa_o}{2} & -i\frac{g_o}{\sqrt2}  & -i\frac{g_o}{\sqrt2}  &  &  &  \\
-i\frac{g_o}{\sqrt2} (n_{g}-n_{e_2}) & -\gamma_o/2 &  & i\Omega &  &  \\
-i\frac{g_o}{\sqrt2} (n_{g}-n_{e_2})  &  & -\gamma_0/2 &  & i\Omega &  \\
 &  i\Omega&  & -\gamma_e/2 &  & -i\frac{g_e}{\sqrt2} (n_{e_1}-n_{e_2}) \\
 &  & i\Omega &  & -\gamma_e/2 & -i\frac{g_e}{\sqrt2} (n_{e_1}-n_{e_2}) \\
 &  &  & -i\frac{g_e}{\sqrt2}  & -i\frac{g_e}{\sqrt2}  & -\frac{\kappa_e}{2} \\
\end{pmatrix}
\\
B &= 
\begin{pmatrix}
\sqrt{\kappa_{o,ext}} & \sqrt{\kappa_{o,int}}  &  &  &  &  &  &  \\
 &  & 0 &  &  &  &  &  \\
 &  &  & 0 &  &  &  &  \\
 &  &  &  & 0 &  &  &  \\
 &  &  &  &  & 0 &  &  \\
 &  &  &  &  &  &  \sqrt{\kappa_{e,ext}} & \sqrt{\kappa_{e,int}}  \\
\end{pmatrix}
\end{split}
\end{equation}
where we have added a factor of $\frac{1}{\sqrt2}$ to $g_o$ and $g_e$ to conserve the total coupling strength ($g_{o,tot}=\sqrt{\sum_i g_{o,i}^2}$). Then, the efficiency is:
\begin{equation}
\eta_{M2O} = |S_{1,7}|^2 = \frac{\kappa_{o,ext}}{\kappa_o}\frac{\kappa_{e,ext}}{\kappa_e} \frac{n_{e1}-n_{e2}}{n_{g}-n_{e2}}\frac{4C_{e}C_{o}C_{a}}{[(1+C_{e})(1+C_{o})+C_{a}]^2},
\end{equation}
where the cooperativities are now ensemble cooperativities, and the coupling strength in the numerator is the sum of squares of the individual coupling strengths. 
Similarly, the added noise referred to the input can be found as:
\begin{equation}
\begin{split}
N_{add,RTI} &= \frac{N_{measured}}{\eta} = N_{wg} + \frac{|S(1,8)|^2}{|S(1,7)|^2} N_{res,int}
\\
&= N_{wg} + \frac{\kappa_{e,int}}{\kappa_{e,ext}} N_{res,int}
\end{split}
\end{equation}
These results are exactly the same as the single atom case due to the symmetry of the S-matrix. Increasing the number of atoms will yield the same results. This indicates that atoms with the same properties (detunings, coupling strengths) can be considered as a single atom with coupling strength equal to ensemble coupling strength to solve the conversion process.

\begin{figure*}
\centering
\includegraphics[width=0.8\linewidth]{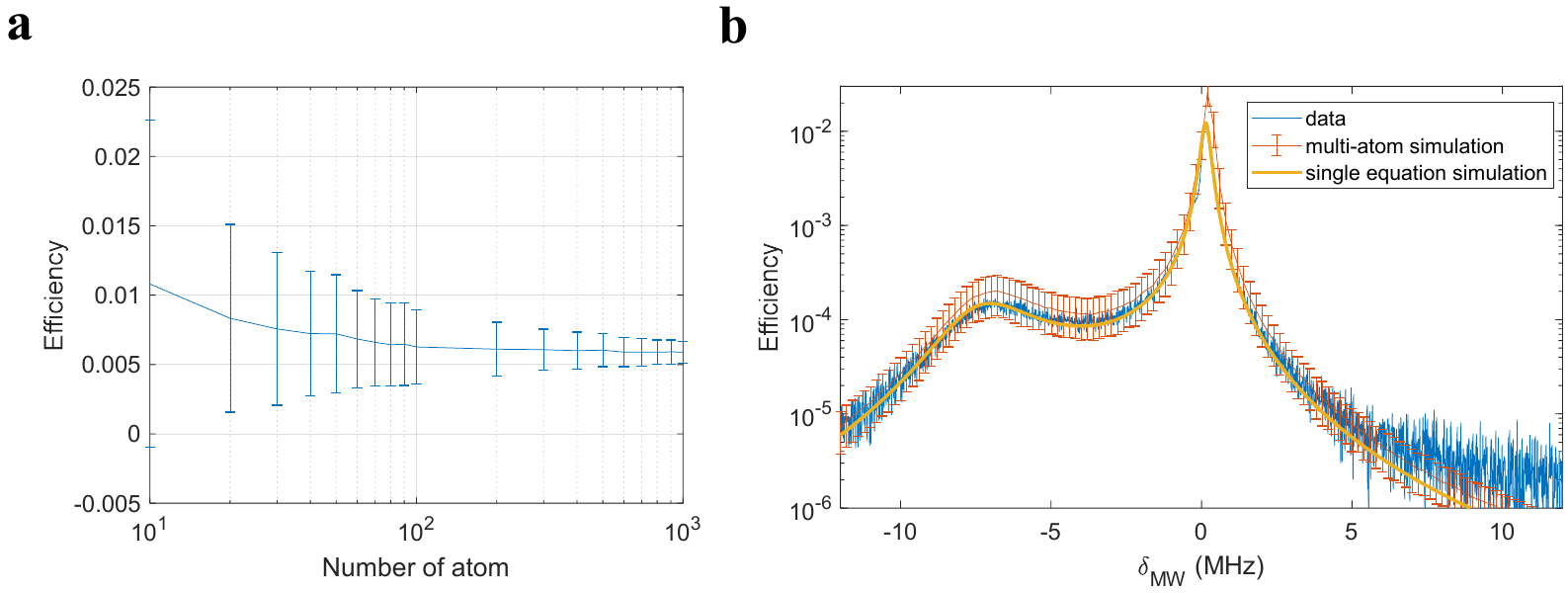}
\caption{\label{S11}
Multi-atom simulations. (a) The convergence test. Parameters used are the same parameters used in the main text Fig.~2b, but with zero spin detuning. Each simulation is repeated 1000 times and the error bars reflect the finite sampling. (b) Comparison between data, multi-atom simulation and the single-atom equation. For the multi-atom simulation, each point consists of 200 atoms and is averaged over 1000 simulations from which error bars are calculated.
}
\end{figure*}

For a large number of atoms with inhomogeneities, deriving an analytical expression becomes intractable, and numerical simulations must be implemented. To do this, we first assign a random optical and microwave detuning to each atom from a Gaussian distribution with width equal to the respective inhomogeneous linewidths. Additionally, random horizontal and vertical coordinates are assigned taking into account a Gaussian distribution based on the optical beam radius. Therefore, from the beam edge to the center, a corresponding optical and microwave coupling strength can be assigned to each sampled atom using the Gaussian beam profile and simulated microwave mode profile. Since we have billions of atoms, we cannot simulate each atom individually. Instead, we group the atoms with similar properties together to reduce the total number of calculated atoms. In doing so, we perform a convergence test to see how many groups of atoms we must sample in order for the result to converge. As shown in Fig.~\ref{S11}a,  around 100 atoms sufficiently capture the system behavior.

Finally, we compare the multi-atom simulation with the single-atom simulation with the same parameters used in the experiments. As shown in Fig.~\ref{S11}b, the results are similar and overlap within a factor of 2. This validates the usage of the simple, single-atom equation as discussed in the main text. However, it is worth noting that these two simulations need not necessarily agree with each other. This is because the single-atom equation uses ensemble cooperativity with inhomogeneous linewidth instead of atomic dephasing in the denominator, which is technically an approximation. Specifically, when the homogeneous linewidth of each individual atom (equivalently, $T_2$) is much narrower than the inhomogeneous linewidth, the result of the single-atom equation deviates from the multi-atom results, due to the fact that the more accurate multi-atom case only addresses a narrow subset of the ensemble. In this case, the denominator of the ensemble cooperativity should be modified to include a contribution from both the homogeneous and inhomogeneous linewidths. In our case, the homogeneous and inhomogeneous linewidths differ by a factor of 40, where the single equation and multi-atom results still agree with each other. 

\section{Resonant $\chi^{(2)}$ derivation}
Here we derive the effective $\chi^{(2)}$ nonlinearity of a three-level REI system. We start with the transduction efficiency of an electro-optical (EO) system in the low cooperativity regime: 
\begin{equation}
\eta = \frac{4C_{eo}}{(1+C_{eo})^2} \sim 4C_{eo}
\end{equation}
Similarly, in the low  cooperativity regime, the atomic system efficiency is
\begin{equation}
\eta = \frac{4 C_e C_o C_a}{((1+C_e)*(1+C_o)+C_a)^2} \sim 4 C_e C_o C_a
\end{equation}
Therefore, an effective $C_{eo} = C_e C_a C_o$ can be found, which leads to an effective coupling rate as
\begin{equation}
g_{eo} = \frac{4 g_{e,tot} g_{o,tot} \Omega_{pump}}{\Gamma_s \Gamma_o}
\end{equation}
The EO coupling rate is related to the material non-linearity as \cite{tsang2010cavity,xu2023efficient}
\begin{equation}
h g_{eo} = \chi^{(2)} V_0 \epsilon_0 c E_{o} E_{p} B_{mw} c
\end{equation}
where $V_0$ is the mode volume, $E_{o}$/$B_{mw}$ is the amplitude of the quantized optical/microwave field, and $E_{p}$ is the pump field strength. Inserting the above equation into the effective EO coupling rate, one finds
\begin{equation}
\chi_{eff}^{(2)} = \frac{h g_{eo}}{V_0 \epsilon_0 c \frac{hg_{o,tot}}{\sqrt{N}\mu_{spin}} \frac{h\Omega}{d_{p}} \frac{hg_{e,tot}}{\sqrt{N}d_{o}}}
= \frac{4}{\epsilon_0 c h^2} \frac{\rho \, d_{p} \, d_{o} \, \mu_{spin}}{\Gamma_s \Gamma_o}
\end{equation}
where $d$ is the electric dipole moment for each optical transition and $\mu$ is the spin magnetic dipole moment. This should be compared to $n^3 r_{ij}$ of EO materials. Also, given the relationship between $g_{eo}$ and $\chi^{(2)}$, one can find an equation linking $\chi^{(2)}$ to efficiency in low cooperativity regime as
\begin{equation}
\eta \sim 4 C_{eo} = \frac{16g_{eo}^2}{\kappa_e \kappa_o} = \frac{4\omega_e \omega_o}{\epsilon_0 V_0} \frac{(\chi^{(2)})^2 P_p}{\kappa_e \kappa_o \kappa_p}
\end{equation}
where $P_p$ is the pump power and $\kappa_p$ is the optical cavity decay rate at the pump frequency. The above equation is meaningful as it indicates the efficiency changes quadratically with the non-linearity $\chi^{(2)}$. Therefore, with a stronger non-linearity, requirements on the quality factors of both microwave and optical resonators are less stringent. 

\section{Microwave mode simulation}
\begin{figure*}
\centering
\includegraphics[width=0.9\linewidth]{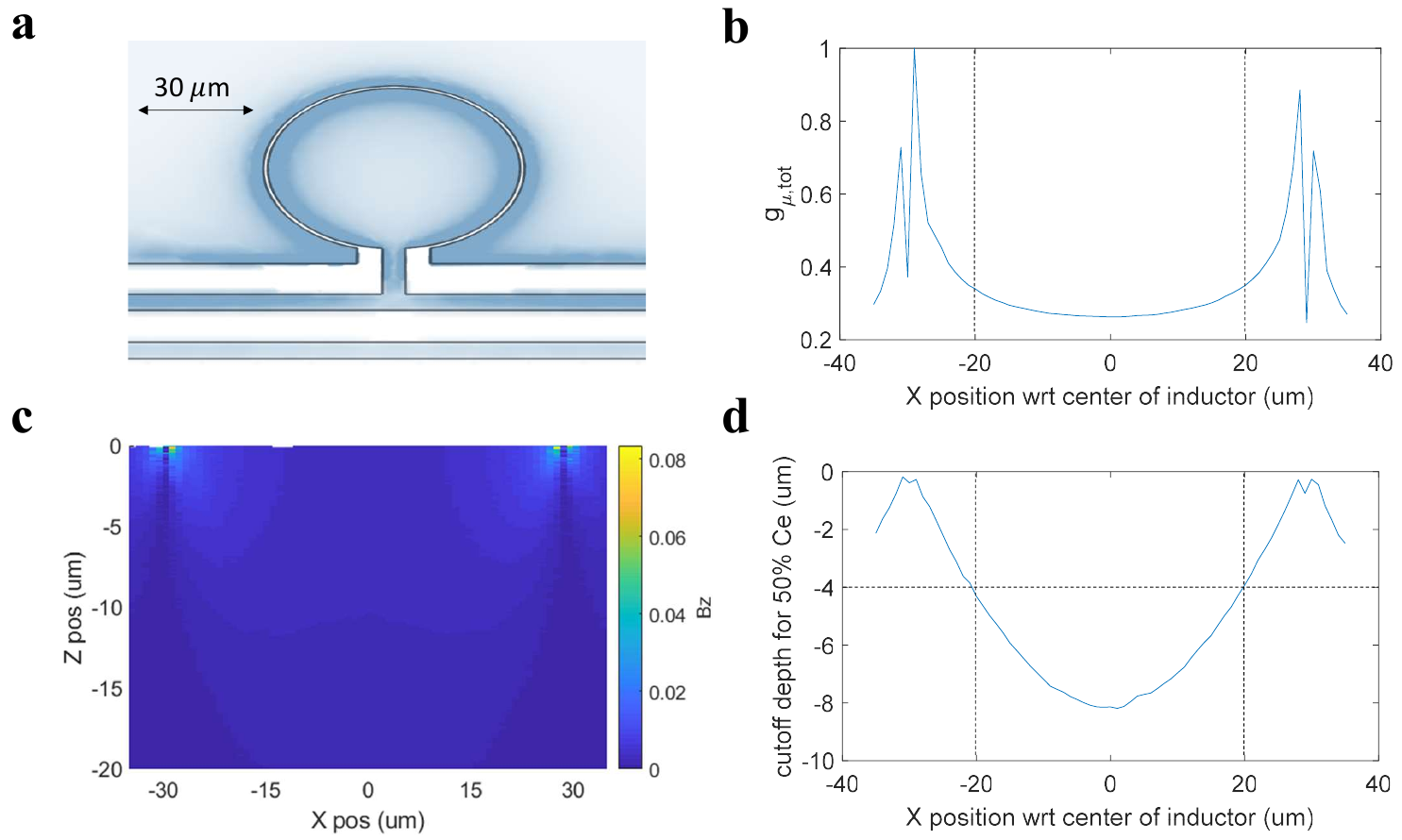}
\caption{\label{S12}
Microwave resonator mode simulation. (a) Top view showing the out-of-plane oscillating magnetic field, which couples to the excited state spin transition. (b) A horizontal cut through the center of the circular inductor, showing the coupling strength at each position on the diameter. (c) A vertical 2D plot showing the penetrating magnetic field strength through the center of the circular inductor. Note that there is no magnetic field along the z-axis right under the inductor as it is purely along the x-axis. (d) The cutoff depth for half of the ensemble cooperativity ($\sim \sum_{i} g_{e, i}^2$).
}
\end{figure*}

The superconducting microwave resonator is made out of a 150 nm niobium film. Each resonator has an interdigitated capacitor formed by 25 capacitor fingers with $\sim$ 570 $\mu$m in length, and an inductor formed by a 1$\mu$m wide line with a 30 $\mu$m bending radius. We simulate the microwave resonator mode profile as shown in Fig.~\ref{S12}a and c. The coupling strength is almost constant within the 20~$\mu$m beam radius (Fig.~\ref{S12}b). To get an estimate of the effective crystal length participating in transduction, we consider a cutoff at half of the cooperativity ($\sim \sum_i g_{e, i}^2$), where the center position shows $\sim$ 8~$\mu$m penetration depth (Fig.~\ref{S12}d). Out of the total magnetic field energy along all directions, $\sim$ 0.27\% of it is confined inside a 20~$\mu$m-radius 8~$\mu$m depth cylinder. In reality, there are complications due to the optical reabsorption of the crystal. Even with 8~$\mu$m thickness, the reabsorption is $\sim35\%$. This roughly balances the 50\% cutoff of the magnetic energy. Without unraveling the balance between the microwave mode cut-off and reabsorption, we find that using this 8~$\mu$m effective length in the simulations predicts the system behaviors well.

\section{Spin-resonator coupling strength simulation}
Considering the ensemble spin-resonator coupling strength
\begin{equation}
g_{e,tot}^2 = \sum_i g_i^2
\end{equation}
where the single spin coupling strength is
\begin{equation}
g_i = \sqrt{\frac{\omega_e \mu_0}{2 \hbar}} \mu_{spin} \frac{1}{\sqrt{V_0}}\frac{B_i}{\sqrt{(|B_i|^2)_{max}}}
\end{equation}
and $\mu_0$ is the vacuum permeability. Putting the above equation into the ensemble equation and converting the discrete sums to integrals,
\begin{equation}
g_{e,tot}^2 = \frac{\omega_e \mu_0}{2 \hbar} \mu_{spin}^2 \rho \eta
\end{equation}
where we define the ratio of the resonator energy confinement to the transduction material $\eta$ as,
\begin{equation}
\eta = \frac{\int_{mat}B^2 dV}{\int_{V_{tot}}|B|^2 dV}
\end{equation}
Apart from material properties, we find that the total ensemble spin-resonator coupling strength is only related to the energy confinement ratio, which has been simulated to be 0.27\% by considering the transduction volume to be a cylinder with radius 20 $\mu$m (equal to beam radius) and length of 8 $\mu$m (from the microwave mode simulations). Therefore, if all of the population is in the excited state, $g_{e,all} = 2.42$ MHz.

Next, we consider the population fraction pumped to the excited state, which depends on the optical pump power. To calculate this, we use the optical Bloch equations in steady state \cite{steck2007quantum}:
\begin{equation}
\rho_{ee}(t\to \infty) = \frac{\Omega^2}{2\gamma_2 \gamma_1} \frac{1}{1+\frac{\Delta^2}{\gamma_2^2}+\frac{\Omega^2}{\gamma_2 \gamma_1}}
\end{equation}
where $\gamma_{1,2}$ are the population decay and coherence dephasing rates, respectively. Therefore, the ensemble excited state population for a given pump power can be calculated as:
\begin{equation}
\rho_{ee} = \frac{\int [\rho_{ee}(\Delta)\frac{1}{\sigma \sqrt{2\pi}} e^{-\frac{1}{2} \frac{\Delta^2}{\sigma^2} }] d\Delta}{\int [\frac{1}{\sigma \sqrt{2\pi}} e^{-\frac{1}{2} \frac{\Delta^2}{\sigma^2} }] d\Delta}
\end{equation}
Using $T_1=267$ $\mu$s measured from a thin crystal (avoiding radiative trapping) \cite{kindem2018characterization} and $T_2=140$ ns measured in Fig.~\ref{ExtFig4}, we get $\rho_{ee} = 0.39$. With a CW pump, the crystal heating will cause the other ground states to be equally populated, and thus the remaining population in the transduction manifold will be 2/3. With this number, we get $g_{e, tot} = 2\pi \times1.24$~MHz under an optical pump with strength $\Omega=2\pi \times1$ MHz, which matches the experimentally extracted value of $g_{e, tot} = 2\pi \times 1.3$~MHz well.

\begin{figure*}
\centering
\includegraphics[width=0.9\linewidth]{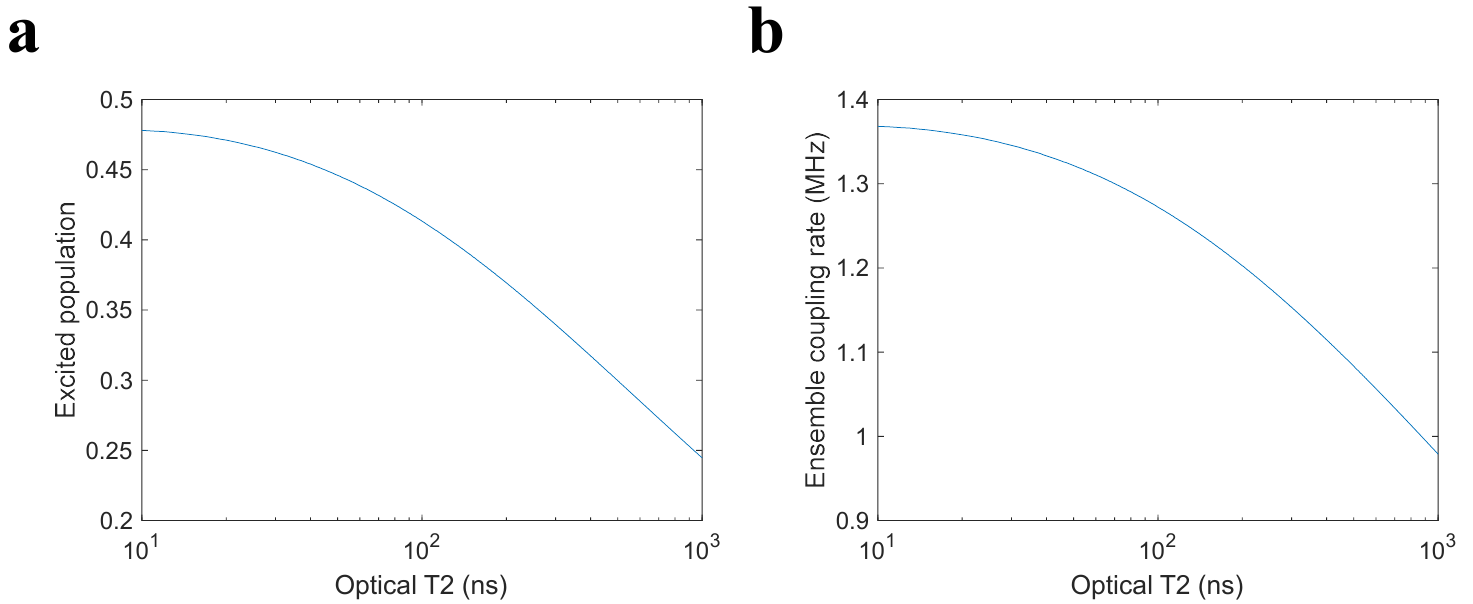}
\caption{\label{S13}
Calculations on excited state population from optical pumping. (a) The excited state population in the ensemble vs the optical $T_2$ of the transition, considering $\Omega=1$ MHz optical pump. (b) The ensemble spin-resonator coupling strength ($g_{e,tot}$) vs the optical $T_2$ of the transition, considering $\Omega=1$ MHz optical pump. 
}
\end{figure*}

\section{Optical cooperativity derivation}
Similar to the microwave resonator-spin coupling, the ensemble total optical coupling strength can be written down as: 
\begin{equation}
\begin{split}
g_{o,tot}^2 &= \frac{\omega_o}{2\hbar\epsilon_r\epsilon_0}d_o^2\rho\eta
\\
\eta &= \frac{\int_{mat}E^2 dV}{\int_{V_{tot}}|E|^2 dV}
\end{split}
\end{equation}
Then, relating the optical dipole moment to the oscillator strength \cite{xie2021characterization}, we connect $g_{o,tot}$ to $\alpha$:
\begin{equation}
\begin{split}
C_o &= \frac{4g_{o,tot}^2}{\kappa_o \Gamma_o} = \frac{4g_{o,tot}^2 \mathcal{F}}{2\pi \cdot FSR \cdot \Gamma_o} = \frac{4g_{o,tot}^2 \mathcal{F} \cdot 2nL}{2\pi c \Gamma_o} 
\\
&= \frac{4\mathcal{F} \cdot 2nL}{2\pi c \Gamma_o} \cdot \frac{\omega_o}{2\hbar\epsilon_r\epsilon_0}d_o^2\rho \eta
\\
d_o^2 &= \frac{\hbar e^2}{2m\omega_o}f
\\
f &= 4\pi\epsilon_0 \frac{mc}{\pi e^2}\frac{1}{\rho}n\pi\alpha_{peak}\frac{\Gamma_o}{2}\frac{1}{2\pi}
\end{split}    
\end{equation}
And finally, we get:
\begin{equation}
C_o = \frac{\alpha_{peak}\lambda Q\eta}{2n \pi} =\frac{\alpha_{peak} L \mathcal{F} \eta}{\pi}
\end{equation}
We note that this result assumes no local correction factor in the oscillator strength equation and a Lorentzian lineshape of the absorption profile. We check this equation by comparing it with the measured $C_o$ in previous works on similar REI systems \cite{lei2023many,rochman2023microwave}. Both of them show a good agreement within a factor of 2. We note that for our transducer, the optical mode is fully inside the YVO$_4$ material, which gives $\eta = 1$.

\section{Noise thermometry}
\subsection{HEMT calibration}
To measure the temperature of the microwave resonator under optical excitation, the microwave output port of the chip (port 3 of the cryogenic circulator) is connected to a low-noise cryogenic amplifier (HEMT) at 4K with superconducting NbTi cables. The HEMT output comes out of the refrigerator and the output power is measured by a spectrum analyzer for CW measurements or mixed down and digitized for pulsed measurements.

In order to back out the power output from the chip, the gain and noise of the entire measurement chain must first be calibrated. Here the gain is composed of the HEMT and any room-temperature amplifiers used, as well as loss from the cables and connections. The noise is dominated by the HEMT as it is held at 4K with a gain of $\sim$40dB, the amplified Johnson noise at the output is much larger than room-temperature. This calibration is done by adding a variable noise source before the HEMT and measuring the output power as a function of this noise. The output power $P_{out}$ can be expressed as \cite{hease2020bidirectional,Xu2020}:
\begin{equation}
\label{poutHEMT}
    P_{out} = \hbar\omega GB(\frac{1}{e^{\frac{\hbar\omega}{k_B T_{N}}}-1}+N_{HEMT}+\frac{1}{2})
\end{equation}
where $\omega$ is the frequency, $G$ is the net gain of the output line, $B$ is the measurement bandwidth, $T_N$ is the temperature of the noise source, and $N_{HEMT}$ is the added noise of the HEMT. Here we use the device as a noise source, and heat the device up with a resisitive heater on the mixing chamber plate. We allow ample time for the entire mixing chamber plate and all components that are thermally lagged to it to thermalize ($\sim$1 hour) before performing measurements. The output power as a function of the thermometer reading is measured and fit to Eq.~\ref{poutHEMT}.
\begin{figure}[h]
    \centering
    \includegraphics[width=0.9\linewidth]{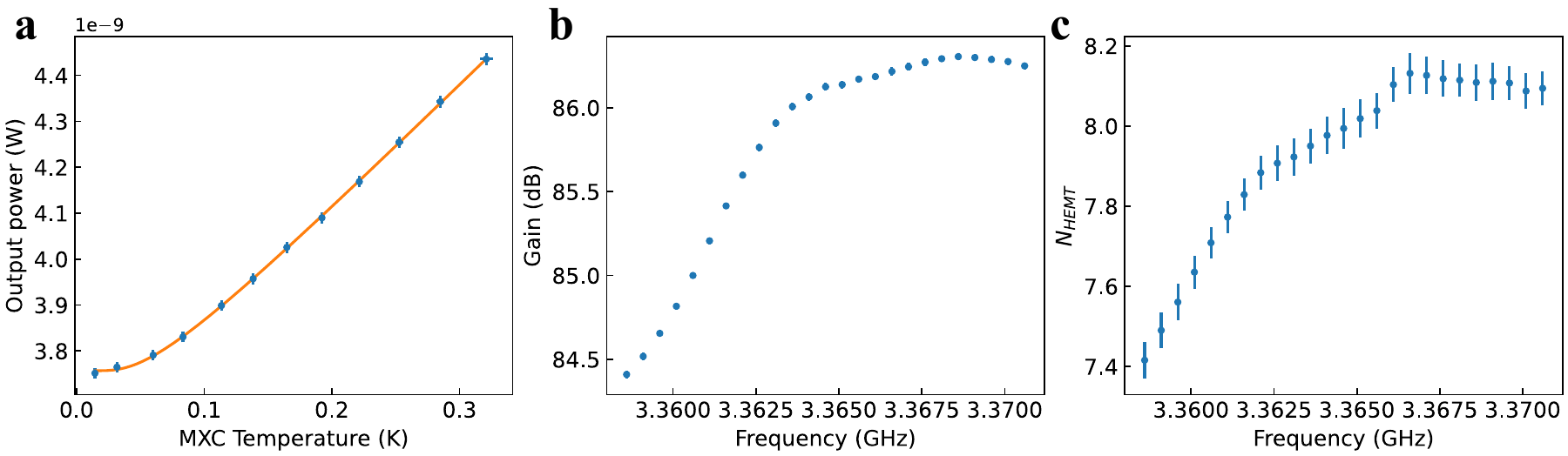}
    \caption{HEMT calibration. (a) The measured output power vs. fridge mixing chamber temperature for a certain frequency, fit to equation\ref{poutHEMT}. (b) Extracted gain of the output line from the fit for different frequencies. (c) Extracted added noise of the HEMT.}
    \label{S14}
\end{figure}

The extracted gain and added noise are used to convert the measured output power to microwave resonator output with optical pulses applied in the main text Fig.~3d and in Fig.~\ref{ExtFig5}.

\subsection{Microwave mode noise}
Applying the calibration to the raw data, we back out the output power directly at the output port of the resonator $S_{out}$. However, the quantities of interest are the resonator photon occupancy $N_{res}$ and waveguide photon occupancy $N_{wg}$. To derive these quantities, we follow the formalism described in \cite{Xu2020}. We start with the driven Tavis-Cummings Hamiltonian, only considering the spin-microwave resonator coupling:
\begin{equation}
    H=\hbar \omega_{a}a^{\dagger}a+\frac{1}{2}\hbar\omega_{at} J^z+\hbar g(a^\dagger J^- +J^+ a)+i\sqrt{\kappa_c}A_{in}(a^\dagger-a)+i\sqrt{\kappa_i}A_{env}(a^\dagger-a)
\end{equation}
where $\omega_a$ is the resonator frequency, $a$ is the resonator mode, $\omega_{at}$ is the spin frequency, $J^z,J^\pm$ are the collective spin operators, $g$ is the spin-resonator coupling strength (although inhomogeneous, later will take the collective coupling strength), $\kappa_c$ is the resonator external coupling rate, $A_{in}$ is the waveguide photon flux, $\kappa_{in}$ is the resonator internal loss rate, and $A_{env}$ is the resonator photon flux. The corresponding equations of motion are:
\begin{equation}
\begin{aligned}
    \dot{a}&=-i\omega_a a-\frac{\kappa}{2}a-igJ^- +\sqrt{\kappa_c}A_{in}+\sqrt{\kappa_i}A_{env} \\
    \dot{J^-}&=-i\omega_{at}J^--\frac{\Gamma}{2}J^- +igJ^za
\end{aligned}
\end{equation}
where we have introduced resonator decay rate $\kappa$ and spin inhomogeneous linewidth $\Gamma$.
Fourier transform the operators to the frequency domain:
\begin{equation}
\begin{aligned}
    a(\omega)&=\frac{\sqrt{\kappa_c}A_{in}+\sqrt{\kappa_i}A_{env}-igJ^-}{\kappa/2-i\Delta}\\
    J^-(\omega)&=\frac{igJ^z(\omega) a(\omega)}{\Gamma/2 - i\Delta_{at}}
\end{aligned}
\end{equation}
where $\Delta=\omega-\omega_a$ and $\Delta_{at}=\omega-\omega_{at}$.
\begin{equation}
\begin{aligned}
    a(\omega)&=\frac{\sqrt{\kappa_c}A_{in}+\sqrt{\kappa_i}A_{env}}{\kappa/2-i\Delta}+\frac{g^2 J^z(\omega)}{(\kappa/2 - i\Delta)(\Gamma/2-i\Delta_{at})} a(\omega)\\
    &=\frac{\sqrt{\kappa_c}A_{in}+\sqrt{\kappa_i}A_{env}}{\kappa/2-i\Delta+C}
\end{aligned}
\end{equation}
where $C=-\frac{g^2J^z(\omega)}{\Gamma/2-i\Delta_{at}}$.
Using input-output formalism $A_{out}=-A_{in}+\sqrt{\kappa_c}a$:
\begin{equation}
\begin{aligned}
    A_{out}(\omega)&=-A_{in}+\sqrt{\kappa_c}(\frac{\sqrt{\kappa_c}A_{in}+\sqrt{\kappa_i}A_{env}}{\kappa/2-i\Delta+C})\\
    &=\frac{\kappa_c -\kappa/2+i\Delta-C}{\kappa/2-i\Delta+C} A_{in} + \frac{\sqrt{\kappa_i\kappa_c}}{\kappa/2-i\Delta+C}A_{env}
\end{aligned}
\end{equation}
The power spectral density for an operator $A$ is
\begin{equation}
    S_{AA}(\omega)=\int_{-\infty}^{\infty} d\tau e^{i\omega\tau} \langle A^\dagger (0)A(\tau) \rangle
\end{equation}
To get $S_{out}(\omega)$, first find the autocorrelation of the output field in the time domain:
\begin{equation}
\begin{aligned}
    \langle A_{out}^\dagger(0) A_{out}(\tau) \rangle &= (\frac{1}{2\pi})^2 \langle \int_{-\infty}^{\infty} d\omega \int_{-\infty}^{\infty} d\omega' e^{-i\omega'\tau} \langle A_{out}^\dagger(\omega) A_{out}(\omega') \rangle\\
    &=(\frac{1}{2\pi})^2 \langle \int_{-\infty}^{\infty} d\omega \int_{-\infty}^{\infty} d\omega' e^{-i\omega'\tau} \frac{(\kappa_c-\kappa_i-2C_R)^2/4 + (\Delta-C_i)^2}{(\kappa/2+C_R)^2+(\Delta-C_i)^2}\langle A_{in}^\dagger (\omega)A_{in}(\omega') \rangle\\&+\frac{\kappa_i\kappa_c}{(\kappa/2+C_R)^2+(\Delta-C_i)^2}\langle A_{env}^\dagger (\omega)A_{env}(\omega') \rangle.
\end{aligned}
\end{equation}
Here we have taken the expectation value of $C$ such that $\langle J^{z\dagger}(\omega) J^z(\omega')\rangle=-N\delta(\omega-\omega')$, where $N$ is the population in the spin ground state. We define $C_R$ and $C_i$ as the real and imaginary parts of $C$, respectively.
The autocorrelators of the spectral densities of the photon fluxes are:
\begin{equation}
\begin{aligned}
    \langle A_{in}^\dagger(\omega) A_{in}(\omega') \rangle &= \bar{N}_{wg}\delta(\omega-\omega')\\
    \langle A_{env}^\dagger(\omega) A_{env}(\omega') \rangle &= \bar{N}_{res}\delta(\omega-\omega')
\end{aligned}
\end{equation}
Plug these in:
\begin{equation}
\begin{aligned}
    \langle A_{out}^\dagger(0) A_{out}(\tau) \rangle &=(\frac{1}{2\pi})^2 \int_{-\infty}^{\infty} d\omega \int_{-\infty}^{\infty} d\omega' e^{-i\omega'\tau} \frac{(\kappa_c-\kappa_i-2C_R)^2/4 + (\Delta-C_i)^2}{(\kappa/2+C_R)^2+(\Delta-C_i)^2}\bar{N}_{wg}\delta(\omega-\omega')\\&+\frac{\kappa_i\kappa_c}{(\kappa/2+C_R)^2+(\Delta-C_i)^2}\bar{N}_{res}\delta(\omega-\omega')\\
    &=\frac{1}{2\pi} \int_{-\infty}^{\infty} d\omega e^{-i\omega\tau}\frac{(\kappa_c-\kappa_i-2C_R)^2/4 + (\Delta-C_i)^2}{(\kappa/2+C_R)^2+(\Delta-C_i)^2}\bar{N}_{wg}+\frac{\kappa_i\kappa_c}{(\kappa/2+C_R)^2+(\Delta-C_i)^2}\bar{N}_{res}
\end{aligned}
\end{equation}
Plug this back into $S_{out}$,
\begin{equation}
\begin{aligned}
    S_{out}(\omega)&=\int_{-\infty}^{\infty} d\tau e^{i\omega\tau}\frac{1}{2\pi}\int_{-\infty}^{\infty} d\omega e^{-i\omega\tau}(\frac{(\kappa_c-\kappa_i-2C_R)^2/4 + (\Delta-C_i)^2}{(\kappa/2+C_R)^2+(\Delta-C_i)^2}\bar{N}_{wg}+\frac{\kappa_i\kappa_c}{(\kappa/2+C_R)^2+(\Delta^2-C_i)}\bar{N}_{res})\\
    &=\frac{(\kappa_c-\kappa_i-2C_R)^2/4 + (\Delta-C_i)^2}{(\kappa/2+C_R)^2+(\Delta-C_i)^2}\bar{N}_{wg}+\frac{\kappa_i\kappa_c}{(\kappa/2+C_R)^2+(\Delta-C_i)^2}\bar{N}_{res}
\end{aligned}
\end{equation}
We can gain some physical intuition by looking at this quantity in the resonant case where $\Delta=\Delta_{at}=0$ and rearranging terms:
\begin{equation}
    S_{out}(\omega=\omega_a=\omega_{at})=\frac{(1-\frac{2\kappa_i}{\kappa}-C_e)^2}{(1+C_e)^2}\bar{N}_{wg}+\frac{4\kappa_i\kappa_c/\kappa^2}{(1+C_e)^2}\bar{N}_{res}
\end{equation}
Where $C_e=\frac{4Ng^2}{\kappa\Gamma}$ is the spin-resonator cooperativity. This shows that the spins will absorb the photons in the waveguide and resonator, and $S_{out}$ represents the output noise power after getting absorbed by the spins.

\section{Spin-phonon coupling rate estimation}
The spin-phonon relaxation process for REIs doped in solids at low temperatures is typically dominated by the direct process, in which phonons resonantly interact with the spins \cite{Abragam2012}. The rate of this process $R$ is \cite{Abragam2012}
\begin{equation}
    R=\frac{1}{\tau_{1}+(1+b)\tau_{ph}}
\end{equation}
where $\tau_{1}$ is the time constant of heat transfer from the spins to the phonons, $\tau_{ph}$ the phonons to the bath, and $b$ is the bottleneck coefficient.

For order of magnitude estimates of these numbers, we first note that $\tau_{1}$ is fast for REIs in solids under 1K, typically on the order of $\mu$s \cite{Abragam2012,Graf1998}. The bottleneck of the thermalization process, as the name suggests, comes from the heat transfer between the phonons and the bath. Specifically, since this is a resonant transfer of heat, we should only consider the phonons resonant within our spin inhomogeneous linewidth, greatly reducing the coupling rate, hence increasing the relaxation time. This is described by the bottleneck coefficient $b$, which is the density of states of the phonon bath resonant with our spins \cite{Abragam2012}:

\begin{equation}
    b=\frac{\rho * 2\pi \nu^3}{2\omega^2\Gamma}tanh^2(\frac{\hbar\omega}{k_B T})\sim 10^8
\end{equation}
where $\rho=4\times10^{24}$~m$^{-3}$ is the REI density, $\nu\sim4$ km/s is the speed of sound in the substrate, and $T$ is the spin temperature. Here we have assumed a spin temperature of $0.5$~K, however this is dependent on how well our sample is thermalized and how much laser power we apply to the chip. Meanwhile the temperature independent parameter $\tau_{ph}$ represents the lifetime of the phonon before it hits a boundary, and can be estimated as $\tau_{ph}\sim L/(2\nu)\sim500$~ns, where $L$ is the length scale of the crystal. Using these estimates, we obtain $R\sim 50$~mHz.

\begin{figure}
\centering
\includegraphics[width=0.5\linewidth]{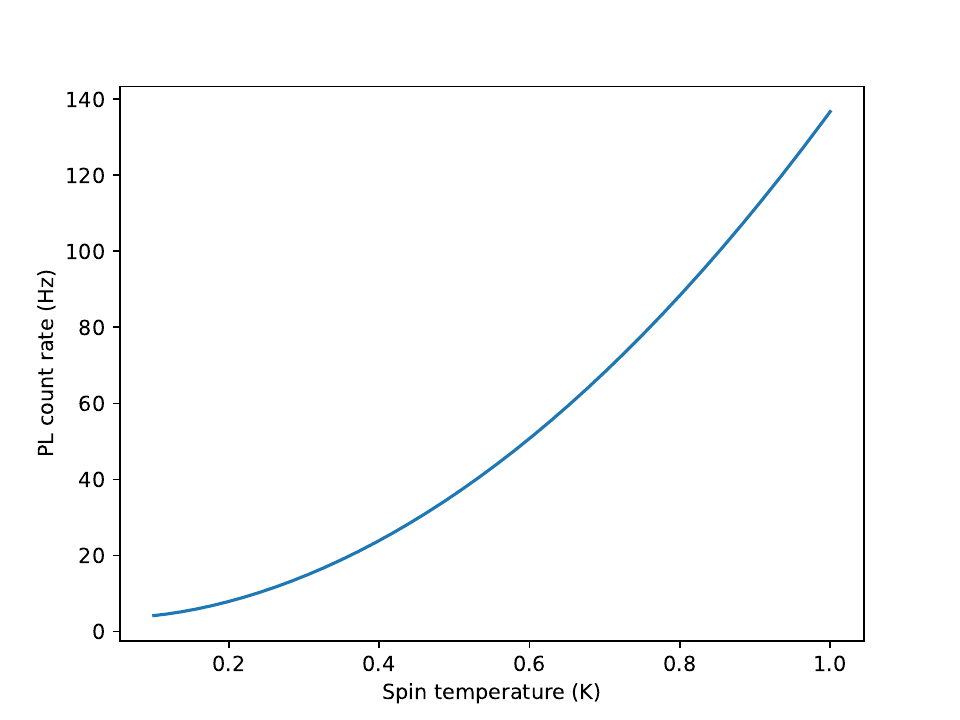}
\caption{\label{S15}
Modeled spin temperature vs. PL count rate expected at the detector, for experimental conditions in Fig.~3a in the main text.
}
\end{figure}

Using this coupling strength, we can also obtain an order-of-magnitude estimate on the expected PL count rate as $\frac{t_{init}R\rho V \rho_{e}}{T_1}$, where we assume spin-phonon coupling occurs during $t_{init}=20$ $\mu$s while the laser is on and the sample is getting heated up, $V\sim\pi (20$ $\mu$m$)^2 \times 8$~$\mu$m is the volume of ions, $\rho_{e}\sim0.05$ is the fraction of ions in the optical excited state after the initialization, and $T_1\sim600$~$\mu$s is the optical lifetime. Under these assumptions, we estimate a few Hz to $\sim100$~Hz PL count rates depending on the spin temperature (Fig.~\ref{S15}). We emphasize again that this is an order-of-magnitude estimate, and we cannot extract a spin temperature from the measured count rate by comparing it to Fig.~\ref{S15}. Rather, this estimation gives some confidence that the PL noise we detect is indeed due to spin-phonon coupling.

\section{Noise PL spectroscopy}
\begin{figure}
\centering
\includegraphics[width=0.80\linewidth]{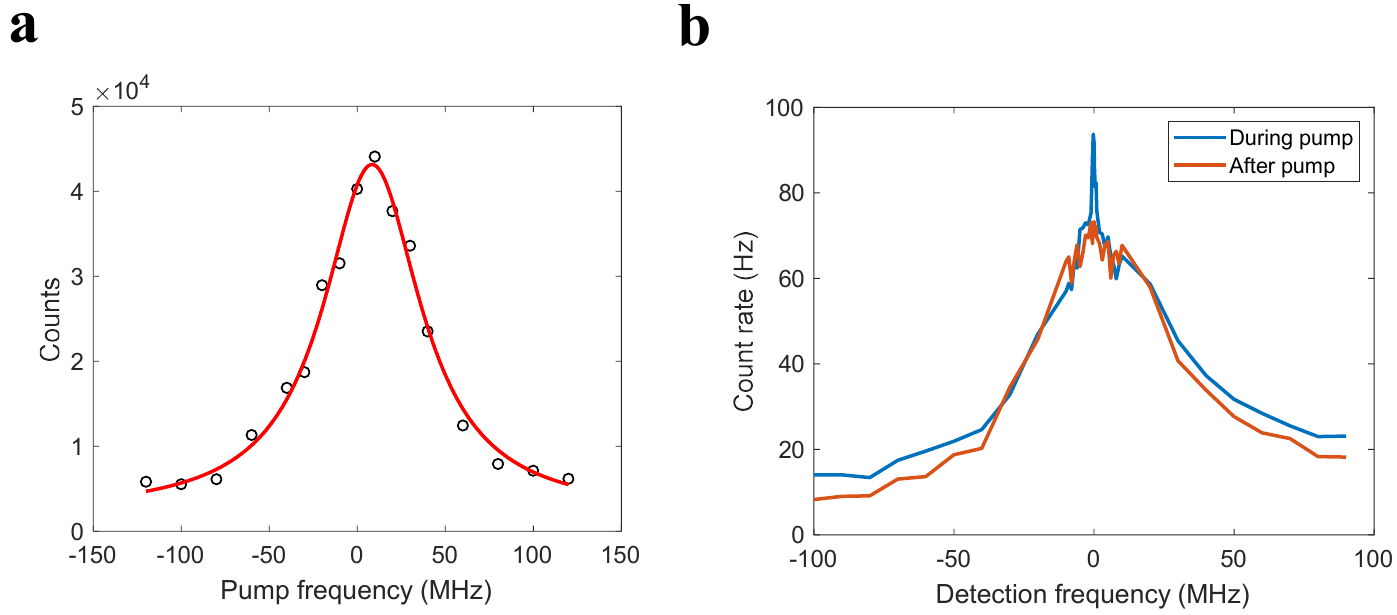}
\caption{\label{S16}
PL measurements with 200 $\mu s$ pulse on and 2 ms pulse off. (a) PL strength vs. excitation frequency, with a fixed detection frequency. (b) PL strength vs. detection frequency, with a fixed excitation frequency. Pump on refers to the emission during the pump, and pump off to the emission right after the pump is turned off.
}
\end{figure}

From the above spin-phonon coupling rate estimation, we believe the PL should be coming from the spin-phonon coupling which transfers population from $|e_1\rangle$ to $|e_2\rangle$. To further confirm this, we pulse the pump with 200~$\mu$s on and 2~ms off, and leave the narrow filters (2~MHz linewidth) at the same detection frequency, but sweep the excitation frequency. If it's from spin-phonon coupling, then the PL strength should have a lineshape centered on the atomic transition frequency of $|g\rangle$ to $|e_1\rangle$, whereas if it's from direct off-resonant excitation of $|e_2\rangle$, the PL strength should be featureless and flat within the experimental scanning range. The results are shown in Fig.~\ref{S16}a, which indeed shows a Lorentzian-shaped PL centered on the transition between $|g\rangle$ and $|e_1\rangle$, with a linewidth of 68~MHz. This is slightly smaller than the optical inhomogeneous linewidth because the spin-phonon coupling rate also depends on the population within that manifold. Moreover, to distinguish the noisy photons from the transduced microwave thermal bath to PL noise photons, we park the pump frequency and scan the filter frequency, effectively tracing out the spectral lineshape of the noise. Then, we separately integrate the PL counts right before turning off the pump (pump on), and right after turning off the pump (pump on). As shown in Fig.~\ref{S16}b, we see the broad PL profile is the same between the pump light on and off cases. However, there is an additional narrow peak only for the emission during the pump. We identify this narrow peak as not PL but rather noisy photons transduced from the microwave resonator bath, as it is much narrower than the PL due to the narrow microwave resonator linewidth. We note that the exact linewidth of this feature is a convolution between the resonator and filter linewidths. In conclusion, we posit that the difference in noise emission during the pump and after the pump is the thermal noise contribution, and confirm that the spectral lineshapes for both this thermal noise and the PL align with our expectations.

\section{Characterization of the second chip used in Fig.~4}
\begin{figure}
\centering
\includegraphics[width=0.70\linewidth]{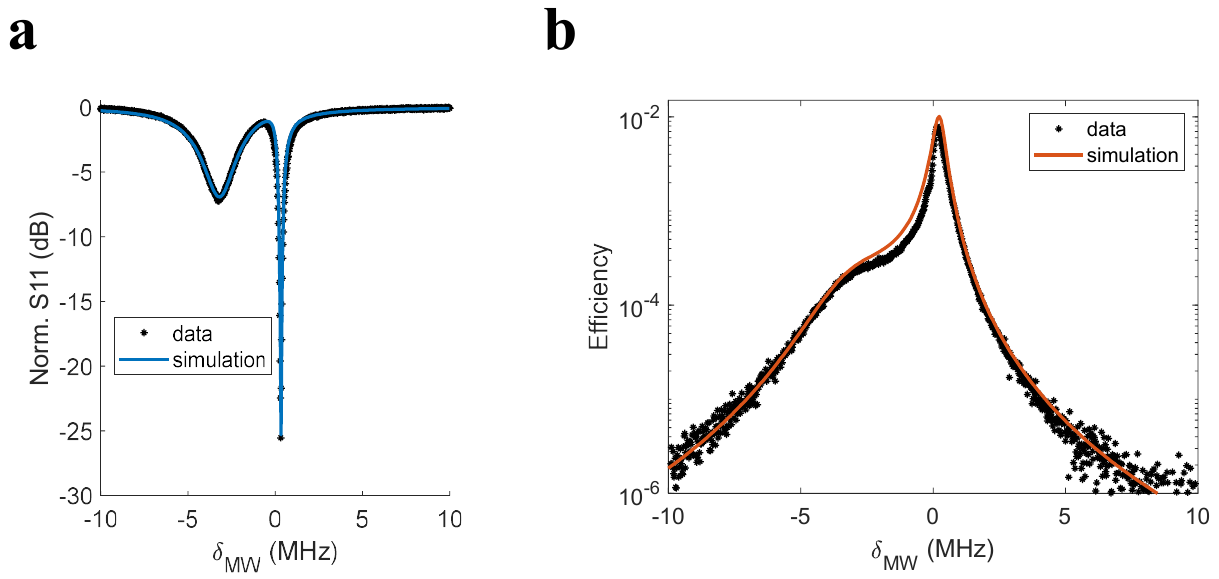}
\caption{\label{S17}
Characterization for the second transducer used in the main text Fig.~4. (a) MW reflection spectrum with the optical pump on, showing the spin resonator coupling. (b) Microwave-to-optical transduction at various input microwave frequencies with the same experimental conditions as (a). 
}
\end{figure}

In Fig.~4 in the main text, we operated two transducers simultaneously and demonstrated photon interference. In Fig.~\ref{S17}, we characterize the second transducer chip not shown in Figs.~1$\sim$3. The microwave resonator has $\kappa_e=2\pi \times4$ MHz with an external coupling ratio of 0.73. In CW mode, with $\sim$1 mW pump power on, the resonator-spin detuning is $\delta_e = -2\pi \times 3$ MHz. The CW efficiency with this pump power is around 1\%, which is similar as the transducer chip showed in the main text.

\let\oldthebibliography=\thebibliography
\let\oldendthebibliography=\endthebibliography
\renewenvironment{thebibliography}[1]{
    \oldthebibliography{#1}
    \setcounter{enumiv}{44}
}{\oldendthebibliography}
\bibliographystyle{naturemag}
\bibliography{refSI}